\documentclass[10pt,journal,a4paper,final,oneside,twocolumn]{IEEEtran}
\usepackage{amsmath,amsfonts}
\usepackage{algorithmic}
\usepackage{algorithm}
\usepackage{array}
\usepackage{textcomp}
\usepackage{stfloats}
\usepackage{url}
\usepackage{verbatim}
\usepackage{graphicx}
\usepackage{cite}
\usepackage{multirow}
\usepackage{multicol}
\usepackage{amssymb}
\usepackage{bbding}
\usepackage{setspace}
\usepackage{booktabs}
\usepackage{tabularx}
\usepackage[numbers,compress]{natbib} 
\usepackage{subfigure}

\usepackage{color}

\title{A Tutorial on Coding Methods for DNA-based Molecular Communications and Storage}

\author{\IEEEauthorblockN{Luping Xiang, \emph{Member, IEEE}, Qiang Liu, \emph{Senior Member, IEEE}, Sirong Chen, \emph{Student~Member, IEEE}, Kang Yan, \emph{Student~Member, IEEE}, Wenfeng Wu, \emph{Student~Member, IEEE}, and Kun Yang, \emph{Fellow, IEEE}}

		\thanks{This work was supported in part by the Natural Science Foundation of China under Grant 62301122 and Grant 62071101; in part by the Fundamental Research Funds for the Central Universities under Grant ZYGX2019J001; in part by the Sichuan Science and Technology Program under Grant 2023NSFSC1375. (\textit{Corresponding Author: Qiang Liu.})}

            \thanks{Luping Xiang, Sirong Chen, Kang Yan and Wenfeng Wu are with the School of Information and Communication Engineering, University of Electronic Science and Technology of China, Chengdu 611731, China, email: luping.xiang@uestc.edu.cn, sirongchen@std.uestc.edu.cn, kangyan@std.uestc.edu.cn, wenfengwu@std.uestc.edu.cn. }
   
    	\thanks{Qiang Liu is with the Yangtze Delta Region Institute (Quzhou), University of Electronic Science and Technology of China, Quzhou, Zhejiang 324000, China, email: liuqiang@uestc.edu.cn.}

        \thanks{Kun Yang is with the School of Computer Science and Electronic Engineering, University of Essex, CO4 3SQ Colchester, U.K, and also with the School of Information and Communication Engineering, University of Electronic Science and Technology of China, Chengdu 611731, China, e-mail: kunyang@essex.ac.uk).}
		 }

\begin{document}

\maketitle

\begin{abstract}
Exponential increase of data has motivated advances of data storage technologies. As a promising storage media, DeoxyriboNucleic Acid (DNA)  storage provides a much higher data density and superior durability, compared with state-of-the-art media. In this paper, we provide a tutorial on  DNA   storage and its role in molecular communications. Firstly, we introduce the fundamentals of DNA-based molecular communications and storage (MCS), discussing the basic process of performing DNA storage in MCS. Furthermore, we provide tutorials on how conventional coding schemes that are used in wireless communications can be applied to DNA-based MCS, along with numerical results. Finally, promising research directions on \textcolor{black}{DNA-based data storage in molecular communications} are introduced and discussed in this paper. 
\end{abstract}

\begin{IEEEkeywords}
DNA, DNA-based storage, DNA-based communications, coding scheme
\end{IEEEkeywords}

\section{Introduction}

The amount of digital data produced has long been outpacing the amount of storage available. Most of the world’s data today is stored on magnetic and optical media. Despite improvements in optical discs, storing a zettabyte of data would still take many millions of units, and use significant physical space. The data explosion has resulted in severe data storage problems, motivating alternative data storage media. DeoxyriboNucleic Acid (DNA) data storage is fast becoming the most cutting-edge solution to the long-term issues expected with current mediums of data storage, such as corruption, degraded hardware, and general matters of sustainability with regards to the manufacture of hundreds of millions of hard drives each year \textcolor{black}{\cite{dong2020dna}}.

DNA has many advantages for storing digital data. For example, the density and durability are superior to those of existing silicon-based storage media. 
 DNA is at least 1000-fold denser than the most compact solid-state hard drive and at least 300-fold more durable than the most stable magnetic tapes. In addition, DNA’s four-letter nucleotide code offers a suitable coding environment that can be leveraged like the binary digital code used by computers and other electronic devices to represent any letter, digit, or other character \textcolor{black}{\cite{ceze2019molecular}}. 

\subsection{Preliminary Exploration of DNA Storage} 

 The idea of using DNA to store information dates back to the 1960s, when Norbert Wiener and Mikhail Neiman first proposed the concept of genetic memory \cite{wiener1964interview,neiman1965molecular}. In 1996, Davis \cite{davis1996microvenus} conducted the first DNA data storage experiment, producing ``Microvenus", a bio-art work made with synthetic DNA and genetically modified bacteria. 
 Inspired by microdots used during World War II, Clelland  \textit{et al.} \cite{clelland1999hiding} concealed encoded information in DNA microdots in 1999, which was the first experiment without living cells. Other researchers have also carried out a number of DNA molecule-based living cell storage works. However, the amount of data stored was rather small due to the limitations of immature DNA synthesis and sequencing technologies at that time.
 
 In 2010s,  advances in biotechnology such as array synthesis, Illumina sequencing, and nanopore sequencing have made it possible to synthesize and read a large number of DNA molecules, catalyzing the development of DNA data storage. A new milestone in DNA storage came in 2012, when Church \textit{et al.} \cite{church2012next} stored 659 KB of data in DNA, where nucleobase ``A" or ``C" was used to represent bit $0$ and ``G" or ``T"  was selected to represent the bit 1, while ensuring that there were no long homopolymers or self-reverse complementary sections. 
 Subsequently, Goldman \textit{et al.} \cite{goldman2013towards}  achieved $739$ KB data storage and introduced error correction to DNA storage, where original data was encoded by Huffman coding, and then converted to nucleobases according to the ternary coding table. Compared with binary coding, ternary coding is capable of controlling the content of ``C" and ``G" nucleobases, avoiding long homopolymers and improving efficiency. 
 However, the ternary coding and quadruple overlapping redundancy fail to fully utilize the storage capacity of DNA, which only achieves a net information density of $0.33$ bits per nucleotide (bits/nt). 
 Bornholt \textit{et al.} \cite{bornholt2016dna} made improvements on the basis of Goldman \textit{et al.}, generating redundant sequences by an exclusive-OR (XOR) operation. 
 However, both studies used specific ways to prevent errors, but neither is an error-correcting code in the true sense.

\subsection{Breakthrough in the Use of Error Correction Codes } 

In order to achieve reliable storage, researchers gradually shifted their focus to implementing error correction coding in DNA storage. In 2015, Grass \textit{et al.} \cite{grass2015robust} introduced finite field-based Reed-Solomon (RS) codes into DNA storage, increasing the net information density to $1.14$ bits/nt. Every two bytes of the original file were mapped to three elements of the finite field $GF(47)$ and each element was mapped to three nucleobases using a codon table. The data matrix was encoded with the outer and inner RS code, so that the insertion, deletion, and substitution errors could be corrected at the same time. Blawat \textit{et al.} \cite{blawat2016forward}  employed the binary Bose–Chaudhuri–Hocquenghem (BCH) codes to protect addresses individually, and RS codes to protect consecutive oligo blocks.
Erlich \textit{et al.} \cite{erlich2017dna} pioneered the use of fountain codes in DNA encoding in $2017$, further increasing the net information density to $1.57$ bit/nt. The fountain codes can generate any number of encoded packets from original $K$ input packets, and the receiver can decode information as long as it receives any $K'$ encoded packets in which $K'$ is slightly larger than $K$. In this case, the encoder simply uses quaternary coding, since the generated DNA sequence that does not meet the biochemical constraints will be discarded, and the next sequence is generated until there are enough sequences. Compared with other codes, the characteristics of fountain codes make them more robust in large-scale storage systems with DNA loss. Two years later, Erlich \textit{et al.} \cite{koch2020dna} further proposed the concept of ``DNA-of-things'' (DoT), putting the DNA encoded by fountain codes into nanometer silica beads, and making a Stanford rabbit containing genetic information with $3$D printing technology, so that data in DNA can be stored stably.

In 2020, Press \textit{et al.} \cite{press2020hedges}  designed a novel inner code scheme, referred to as Hash Encoded, Decoded by Greedy Exhaustive Search (HEDGES), which is capable of correcting insertion, deletion and substitution errors within the DNA sequence. The algorithm works in two steps: (1) perform a hash operation on the strand ID, bit index, and several previous bits corresponding to each binary bit of the sequence to be encoded; (2) add the calculated value to the current bit value, and convert it to a nucleobase after performing mod $4$. 
In the decoding process, a greedy search on an expanding tree of hypotheses is performed to locate and correct the error.
Song \textit{et al.} \cite{song2021robust} designed a sequence reconstruction algorithm based on the de Bruijn graph \cite{zerbino2008velvet} to solve DNA breakages and rearrangements. The algorithm combines greedy path search and cyclic redundancy check (CRC) codes to achieve efficient de novo strand assembly of broken DNA fragments, which leads to the long-term reliability of DNA storage. After the reconstruction of DNA fragments, the original data can be perfectly restored by using fountain codes to restore a small amount of lost sequences.
Chen \textit{et al.} \cite{chen2021artificial} designed and synthesized an artificial yeast chromosome for data storage based on LDPC and watermark code. The LDPC-encoded data was interleaved and then added to a pseudo-random sequence (watermark). Pseudo-random sequences were employed to identify and locate insertions and deletions, similar to the function of hidden hints in puzzles. 
This scheme combines the synchronization code of positioning insertion and deletion with traditional error correction codes, achieving better error correction performance at the cost of higher decoding complexity when compared with RS codes.
In 2022, Pan \textit{et al.} \cite{pan2022rewritable} developed a two-dimensional DNA data storage system termed 2DDNA that records information in both the sequence and the backbone structure of DNA and performs joint data encoding, decoding, and processing. Their approach combined machine learning (ML) and computer vision (CV) techniques for image reconstruction and enhancement. For some images with highly granular details, they also proposed error protection methods based on LDPC.

\subsection{Other Advances in DNA Storage } 

DNA synthesis and sequencing naturally generate multiple copies, so multiple sequence alignment (MSA) algorithms can be used. The DNA storage systems of Antkowiak \textit{et al.} \cite{antkowiak2020low} and Yazidi \textit{et al.} \cite{yazdi2017portable} employed a strategy of first inferring approximately consensus sequences via MSA, and then correcting remaining errors via additional code constraints. Xie \textit{et al.} \cite{xie2023study} conduct a simulation study on the error correction capability of a typical MSA algorithm, MAFFT. Unlike conventional error-correcting codes (ECC), MSA can resolve IDS errors without adding logical redundancy or estimating any model parameters, provided sufficient sequencing depth is available.

In addition to optimizing coding algorithms, DNA storage also made progress in other areas as given in reference \cite{doricchi2022emerging}, such as expanding the nucleobase system, enabling random access and achieving more convenient DNA synthesis and sequencing.

By using additional artificial nucleobases or degenerate bases, nucleobases may be expanded to increase the storage density. In 2019, a DNA system constructed from eight nucleotides ``letters" was designed in \cite{hoshika2019hachimoji}. In the same period, Anavy \textit{et al.} \cite{anavy2018improved,anavy2019data} employed composite DNA letters in the coding process. Composite DNA letters are degenerate bases, representations of a position in a sequence that contain four kinds of nucleobases mixed in a special predetermined ratio. 

Earlier DNA storage methods were required to decode the entire file to read a piece of data, which significantly increased time and economic costs. Therefore, many studies on random access have emerged, for example, random access to specific files using PCR \cite{tabatabaei2015rewritable,organick2018random,organick2020probing}, or selection of specific files using fluorescence-activated sorting (FAS) through the reversible binding of fluorescent probe molecules to the file barcodes \cite{banal2021random}. 

The development of DNA-based data storage also relies on the optimization of synthesis and sequencing technologies to increase the speed of reading/writing information and reduce costs. Relevant studies include realizing random-access and error-free data recovery from low-cost and portable devices \cite{yazdi2017portable}, and integration of DNA synthesis and sequencing on miniaturized SlipChip device \cite{xu2021electrochemical}. Microsoft and the University of Washington have built an automated end-to-end DNA data storage device that can automatically write, store, and read data \cite{takahashi2019demonstration}. 

\begin{figure*}[htbp]
	\centering
	\includegraphics[width=6in]{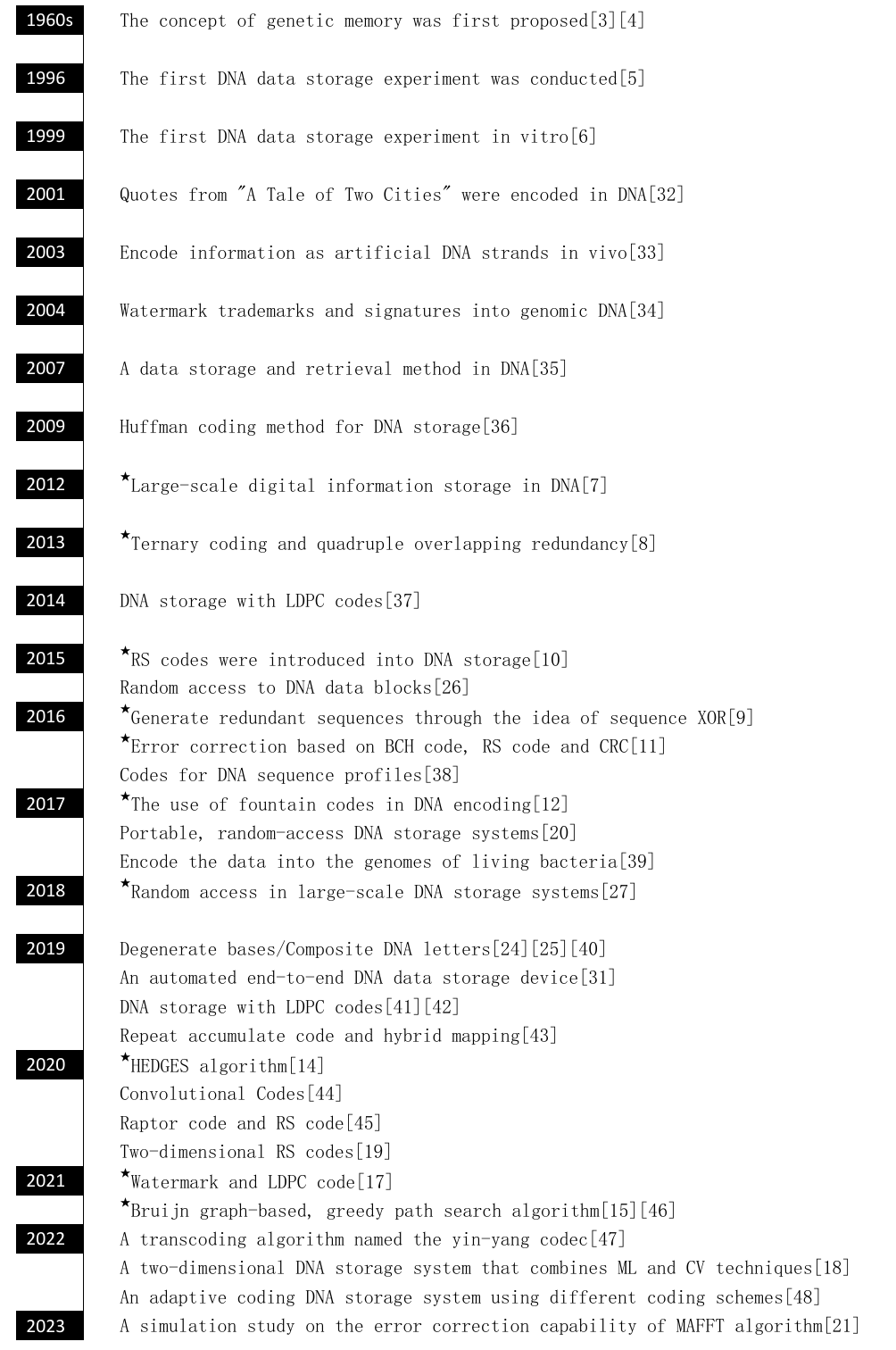}
	\caption{Literature related to DNA data storage (Pentagrams indicate representative articles related to DNA storage or articles that use a common DNA coding method)
	\cite{
	wiener1964interview, neiman1965molecular,
	davis1996microvenus, 
	clelland1999hiding,
	bancroft2001long,
	wong2003organic,
	arita2004secret,
	yachie2007alignment,
	ailenberg2009improved,
	church2012next,
	goldman2013towards,
	yim2014essential,
	grass2015robust,
	tabatabaei2015rewritable,
	bornholt2016dna,
	blawat2016forward,
	kiah2016codes,
	erlich2017dna,
	yazdi2017portable,
	shipman2017crispr,
	organick2018random,
	anavy2018improved,anavy2019data,choi2019high,
	takahashi2019demonstration,
	chandak2019improved,deng2019optimized,
	wang2019high,
	press2020hedges,
	chandak2020overcoming,
	shufang2020dna,
    antkowiak2020low,
	chen2021artificial,
	song2021robust,song2022robust,
    ping2022towards,
    pan2022rewritable,
    cao2022adaptive,
    xie2023study}}
	\label{LiteratureRelatedToDNADataStorage}
\end{figure*}

\subsection{Contributions \& Structure} 

An ideal storage system should have the characteristics of low read/write delay, high throughput, and high reliability, and the working efficiency required to meet the needs of equipment computing or communication. Nowadays, widely used information storage technologies, such as portable hard disks, USB flash memory or integrated circuits, have defects such as low stability and large environmental pollution. DNA molecular communication is the key technology to solve such thorny problems since it has the advantages of high stability, high capacity, environmental protection, etc.
Reference \cite{bi2021survey} facilitated the development and implementation of interdisciplinary solutions in the field of molecular communications by mapping phenomena, contributions, and problems into hierarchies. References \cite{yazdi2015dna} introduced the overall process of DNA storage, including storage system design methods and its DNA synthesis, sequencing, editing, etc., while providing classic DNA storage coding principles and simple examples. Although there are many tutorials covering DNA storage, errors that occur in DNA storage systems are still a problem that needs to be detailed. Therefore, this article focuses on coding schemes for correcting errors in DNA storage. Hopefully, this paper will spark more research in this promising field of DNA-based molecular communications and storage (MCS).
We provide a brief tutorial on coding methods for DNA-based MCS. Our contributions are summarized as follows.

\begin{itemize}
    \item Firstly, we provide an overview on the development of DNA storage, \textcolor{black}{from early beginnings to the current state-of-the-art. Key milestones and breakthroughs in DNA coding methods are highlighted and other advances that could facilitate DNA storage are outlined.}
    \item Secondly, we provide a tutorial on the DNA-based MCS. \textcolor{black}{We delve into the relevant fundamental concepts, including DNA characteristics, the basic process of MCS, types of DNA errors, and other constraints to consider when coding.}
    \item Thirdly, we provide a tutorial on DNA storage coding, \textcolor{black}{including basic encoding and decoding algorithms for several classic erasure codes and error correction codes, toy examples, and performance simulations to give readers a comprehensive understanding of its method and performance.}
\end{itemize}
The rest of this work is structured as follows. Section II provides fundamentals of DNA-based MCS. Section III reviews typical erasure codes and their applications in DNA storage, while Section IV explains error correction codes for nucleotide substitution. Section V points out the bottlenecks of existing storage coding methods and future research directions.

\section{Fundamentals of DNA-based Molecular Communications and Storage}
Before diving into the details of DNA storage, we first briefly review the basics of DNA-based MCS.
In DNA-based MCS,  DNA is employed as a carrier for information storage and transmission. 
In this section, we first discuss the characteristics of DNA in Section \ref{sec.CharacteristicsDNA}. DNA-based communications and storage are reviewed in Sections \ref{sec.molecular} and \ref{sec.storage}, respectively.


\subsection{Characteristics of DNA}
\label{sec.CharacteristicsDNA}

Composed of deoxynucleotides, DNA exists in biological cells and stores individual genetic information. Each deoxynucleotide in the DNA chain contains four choices, namely: adenine (A), cytosine (C), guanine (G), or thymine (T). The probability of these four nucleotides appearing at a base pair position in the DNA single strand is equal, and each DNA single strand sequence differs from one another, e.g. ATTCGGC.

\begin{figure*}[!t]
	\centering
	\includegraphics[width=5in]{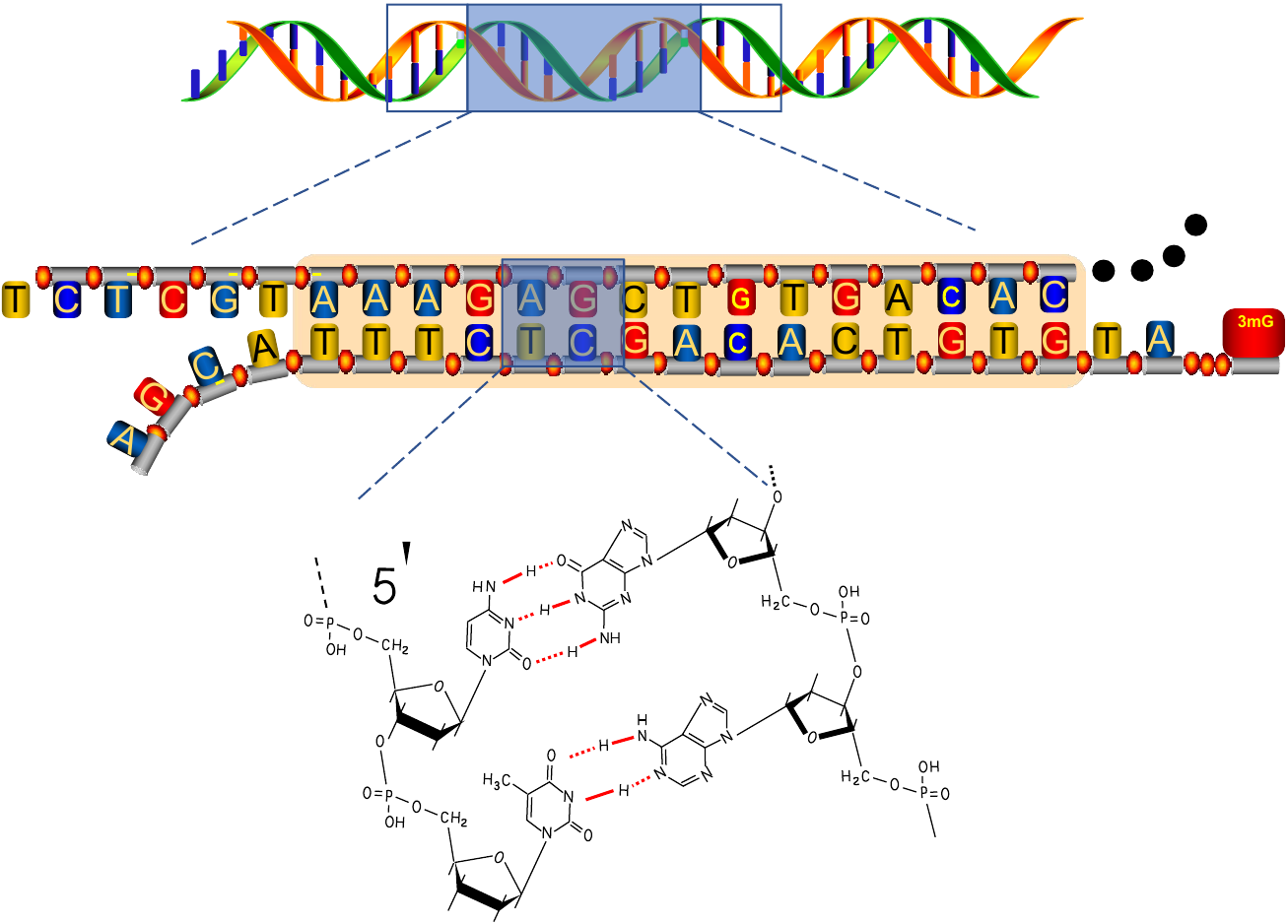}
	\caption{The double helix structure of DNA.}
	\label{DoubleHelixStructureOfDNA}
\end{figure*}

One of the most important properties of DNA is the spatial double-helix structure, as shown in Fig. \ref{DoubleHelixStructureOfDNA}, where a DNA consists of two antiparallel strands of deoxynucleotides. The deoxynucleotides on the chain are connected with their complementation by hydrogen bond, and the connection mode follows the strict law of base complementary pairing, that is: A and T, G and C. Hence, the base pair sequences of the two single DNA strands are completely complementary. 
Since the biological structure determines the properties, the complex double helix structure makes DNA molecules have the following distinct characteristics: 

\begin{itemize}

\item{Stability: The DNA double-strand formed by phosphate-linked nucleotides is stable. Additionally, the A-T pairing forms two hydrogen bonds, while the C-G pairing forms three hydrogen bonds,  ensuring the stability of the double helix structure. 
}

\item{Diversity: The length and base pair sequence of DNA molecules are variable. Since each base pair has four choices,  the DNA molecular sequence has rich change space, which makes DNA molecules an ideal high-density information storage medium.}

\item{Specificity: Because there are four choices for each base position in the DNA chain, the base pair sequence of each DNA molecule represents unique biological information, which becomes the unique information identification of the DNA molecule as well.}

\end{itemize}

Since the DNA information is contained in the size and base sequence of DNA strands, how to efficiently and accurately translate the information contained in DNA becomes a key issue for both DNA storage and communications. 
Hence, we provide a brief introduction of \textit{writing} and  \textit{reading for} DNA for readers who are less familiar with this process.


\textcolor{black}{The \textit{writing} process of DNA information is actually the synthesis process of DNA strand, and the most widely employed method is the phosphoramidite chemistry method proposed by Marvin H. Caruthers \cite{matteucci1981synthesis,beaucage1981deoxynucleoside}. This method is characterized by high efficiency, high speed and high stability in initial reactants \cite{caruthers2013chemical}. 
In addition, new technologies developed in recent years, such as enzymatic synthesis, have also advanced faster and more efficient DNA synthesis.}

\textcolor{black}{The \textit{reading} process refers to the sequencing of the DNA. 
The traditional sequencing technologies include Sanger sequencing \cite{sanger1977dna} and chemical degradation of Maxam and Gilbert \cite{maxam1977new}. 
Now, the most popular method is Next Generation Sequencing (NGS) \cite{goodwin2016coming}, a typical representative of which is the sequencing platform of Illumina. 
Nanopore sequencing \cite{deamer2016three}, one of the third-generation sequencing technologies, is also a promising and potential detection method. Due to the volume advantage of nanopores, it can be used to complete the task of high-throughput DNA sequencing. }

\subsection{DNA-based Molecular Communication}
\label{sec.molecular}

To overcome serious diffraction of electromagnetic waves in a liquid environment, molecular communications could be adopted in water environment detection, directional communication, drug-targeted transportation, etc. As an information transmission medium, DNA has the characteristics of high density and stability, which makes up for the low efficiency of molecular communication to a certain extent. 

The DNA-based molecular communication process includes encoding, emission, transmission, reception, and decoding \cite{sun2019channel}, and the design of each step needs to fully consider the characteristics and performance of DNA and nano-machine.

\begin{figure*}[!t]
	\centering
	\includegraphics[width=5in]{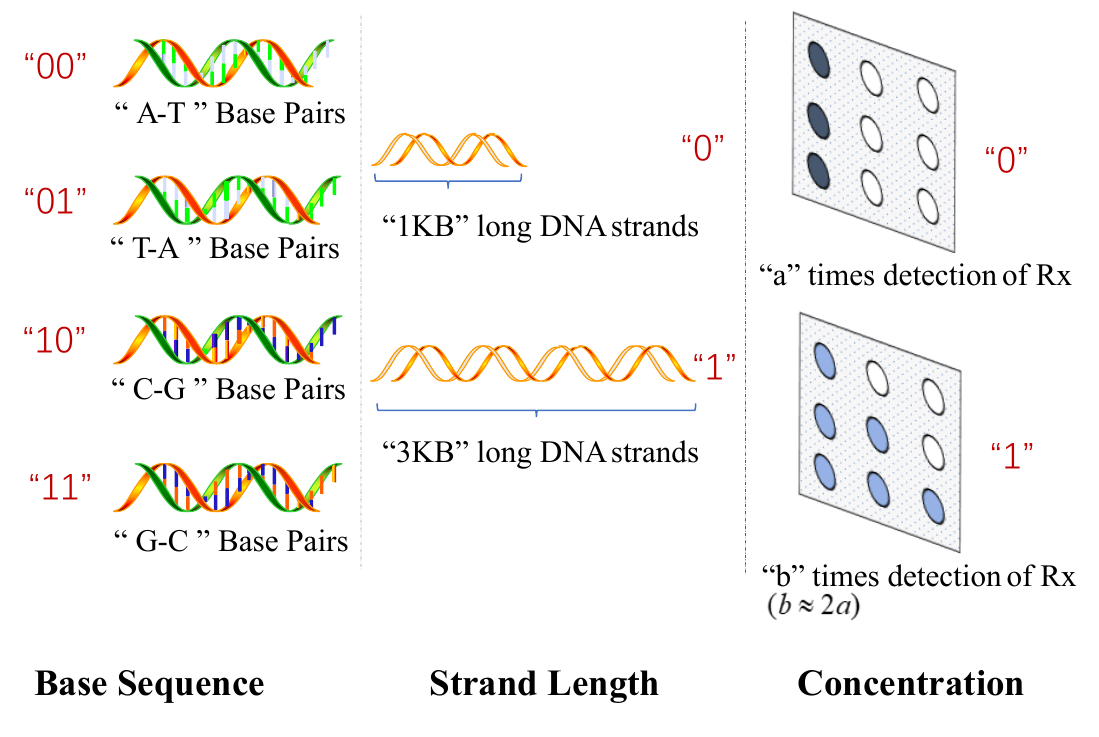}
	\caption{The encoding mechanism of DNA molecules.}
	\label{EncodingMechanismOfDNAMolecules}
\end{figure*}

\textbf{Encoding}: Encoding is the process of converting ``$01$” bit stream into biological information embedded inside DNA \cite{bornholt2017toward}. There are multiple information expression forms such as length and base pair sequence inside the DNA strand. When the DNA strand is released from the transmitter as a molecule, the releasing time and molecular concentration act as additional information expression vectors. 
As shown in Fig. \ref{EncodingMechanismOfDNAMolecules}, coding methods for DNA-based molecular communication are usually divided into three categories: (1) Base pairs sequencing encoding \cite{shah2017molecular,sun2019channel}. Use different bases to represent information. Each nucleotide has four options of A, T, C, and G, and can contain up to 2 bits of information. (2) Length encoding \cite{bilgin2018dna,xie2021channel}. Assuming that there are only two lengths of DNA molecules, we may express the long chain as ``$1$" while the short chain as ``$0$”. (3) Concentration encoding. The high-concentration information molecule recognized by the receiver could be represented as ``$1$” and the low concentration as ``$0$”. In addition, the release time and rotation direction of molecules could also be converted as information bits similarly. At present, DNA molecular communication is still in its infancy, and the coding involved in the system is more similar to the concept of modulation in wireless communication, rather than complex error-correcting codes. More information related to the coding of DNA-based communication can be obtained by referring to the literature on molecular communication \cite{kuscu2019transmitter,hofmann2023coding,farsad2016comprehensive}.

\textbf{Emission}: Emission is the process by which the transmitter releases a specific number of DNA molecules at a fixed time, assuming that the transceivers are synchronized. In order to realize these functions, the transmitter needs to meet the following design requirements. First, the transmitter design volume should be minimized to nanoscale/microscale as far as possible \cite{kuscu2019transmitter}. Since embedding the battery to nanomachines seems to be infeasible, realizing the recycling of energy is also the function required by the transmitter \cite{kuscu2019transmitter}, so do the sustainability of molecules \cite{huang2017capacity}. In order to generate specific kinds of DNA molecules, the transmitter needs to have basic base recognition and editing capabilities. Since the DNA molecular communication system needs to be deployed in the body's fluid environment, the transmitter needs to be biocompatible, e.g., adopting the genetically modified bacteria and cell structure.

\textbf{Transmission}: Once the DNA molecules are released into the environment, they begin the transmission process. In an ideal liquid environment, e.g., body fluid environment, the motion of molecules in it can be described as Brownian motion. The movement of molecules is driven by collisions between solute molecules or between solute and solvent molecules. The probability that the molecule will reach the point with a distance of $d$ after a time $t$ from the origin can be described as \cite{noel2015joint,nakano2012channel}

\begin{equation}
	\label{deqn_ex1a}
	f(d, t)=\frac{d}{\sqrt{4 \pi D t^{3}}} e^{-\frac{d^{2}}{4 D t}},
\end{equation}
where $D$ is the diffusion coefficient related to the molecular length. When the transmitter emits a large number of molecules, the motion of individual molecules is random and the motion of collective molecules is chemotactic.

Another typical transmission medium is blood vessels, where the movement of molecules is controlled by three factors: blood drift, free diffusion, and intermolecular collisions. Molecular motion in blood vessels can also be characterized as \cite{chen2020parameter}

\begin{equation}
	\label{deqn_ex1a}
	\begin{aligned}
		f(t, \mu, \lambda)&=\left(\frac{\lambda}{2 \pi t^{3}}\right)^{\frac{1}{2}} \exp \left[\frac{-\lambda(t-\mu)^{2}}{2 \mu^{2} t}\right], \\
		\mu&=\frac{d}{v}, \\
		\lambda&=\frac{d^{2}}{2 D},
	\end{aligned}
\end{equation}
where $\mu$ represents the time taken for blood flow to flow through a specified distance, and  $\lambda$ represents the influence of diffusion motion in molecular motion. When judging whether diffusion or advection is dominant in molecular propagation, the relationship between  $\mu$ and $\lambda$  can be compared \cite{wicke2018modeling}. When  $\mu$  is much larger than $\lambda$, diffusion is the main propagation scenario, and vise versa. By adjusting different values of $\mu$ and $\lambda$ , we can simulate all kinds of blood vessels from the aorta to the capillaries.

In diffusion-based DNA communication, system performance will be negatively affected by inter-symbol interference (ISI). Time-slotted communication schemes are often considered, where DNA oligos are released at the beginning of each time slot. To avoid ISI, molecules of the same type should be released at time points far enough apart, which limits the communication rate \cite{arjmandi2016isi}. Due to the presence of molecules released from previous time slots in the environment, Rx needs to detect that the concentration of molecules reaches a certain threshold. Thresholds for different DNA oligos need to be calculated accordingly and optimized to minimize the impact of ISI. For length encoding, the diffusion constant of the DNA oligo is also related to its length, so the threshold for detecting different types of DNA oligos is also different \cite{bilgin2018dna}.

\textbf{Reception}: Reception refers to the process by which the receiver captures DNA molecules at a specific time and completes the information collection and processing of target molecules. Similar to the emission process, the design of the receiver should satisfy the following requirements: the volume needs to be limited to the nanoscale and it is biocompatible. It needs basic DNA base recognition ability to operate the molecules arrived. Since the arrival of molecules at the receiving end is a continuous process, the receiver needs to work continuously within a certain number of slots to monitor the arriving DNA molecules. 

\textbf{Decoding}: The decoding process is reversed to the encoding discussed above, which converts the biological information of DNA into bit information. For example, a nanopore is an electronic device which detects the molecular information concurrently. When DNA molecules spin through, it can convert the length information of the DNA strand into current blocking time, and the DNA sequence can also be obtained \cite{bilgin2018dna}. Although DNA molecules have a stable double helix structure, nucleotides may be mutated into other types due to the influence of electromagnetic radiation and oxidants, which leads to mutation and wrong assembly of base sequences \cite{sun2019channel}.

The main differences between DNA-based molecular communication (MC) and traditional communication systems using electromagnetic (EM) signals are as follows: (1) Diffusion-based propagation in DNA-based MC will result in a relatively large delay spread, and more consideration will be given to release time slots and time delays; (2) Symmetrical channels are used in many cases in EM communication, while MC is an asymmetrical channel \cite{hofmann2023coding}; (3) Due to system complexity, many complex codes in EM communication have not been introduced in MC for the time being \cite{liu2021dna}.

At present, the research on DNA-based molecular communication systems is mainly at the theoretical level. When considering system implementation, the following aspects need to be considered: (1) Technical bottleneck. The research scene of DNA-based molecular communication focuses on the nano level, and the transceiver is designed at the nano level to achieve high-precision control of the transmission and reception of DNA molecules. Therefore, the technological progress based on nanomachines is of great significance \cite{prakash2007microfluidic,fuerstman2007coding}; (2) Transmission performance. DNA-based molecular communication is a multi-factor control system. In order to realize high-capacity transmission, it is necessary to optimize the transmission distance, time-slot length, the number of nanopores, and other factors \cite{liu2021dna}; (3) Synthesis expenditure. In the large-scale application of DNA-based molecular communication systems, the actual cost estimation is inevitable. According to the development level of gene editing and sequencing technology, the cost of synthesizing DNA strands is about \$12400 / MB \cite{goldman2013towards}. The high cost of DNA strand synthesis and long reading and writing delays hinder the application of DNA vectors in molecular communication systems.  

\subsection{DNA-based Storage}
\label{sec.storage}


Compared with traditional  hard-disk or flash-memory storage technology, DNA storage has the following characteristics: 

\textbf{High stability}: 
The unique double helix structure of DNA is very stable and does not easily react with the outside world. At the same time, due to the fixed pairing and replication rules of nucleotides, the probability of errors in the process of DNA information transcription and replication is very low.

\textbf{High information capacity}: DNA is an ideal information storage medium because of its high information storage density. A DNA unit can store 200PB information, equivalent to the capacity of 100000 portable hard disks \cite{liu2021dna}. The data with picture and video information can be inserted into the modified bacterial gene. With the independent proliferation of bacteria, the automatic replication of stored information can be achieved. 

\textbf{Long storage lifetime}: Theoretically, DNA can be stored for thousands of years in dark, dry, and cold environments \cite{goldman2013towards}. Moreover, if DNA is dehydrated, its biological structure can be maintained for a long time, and it can be reactivated when information needs to be extracted \cite{bonnet2010chain}. 
 \begin{figure*}[!t]
 	\centering
 	\includegraphics[width=6in]{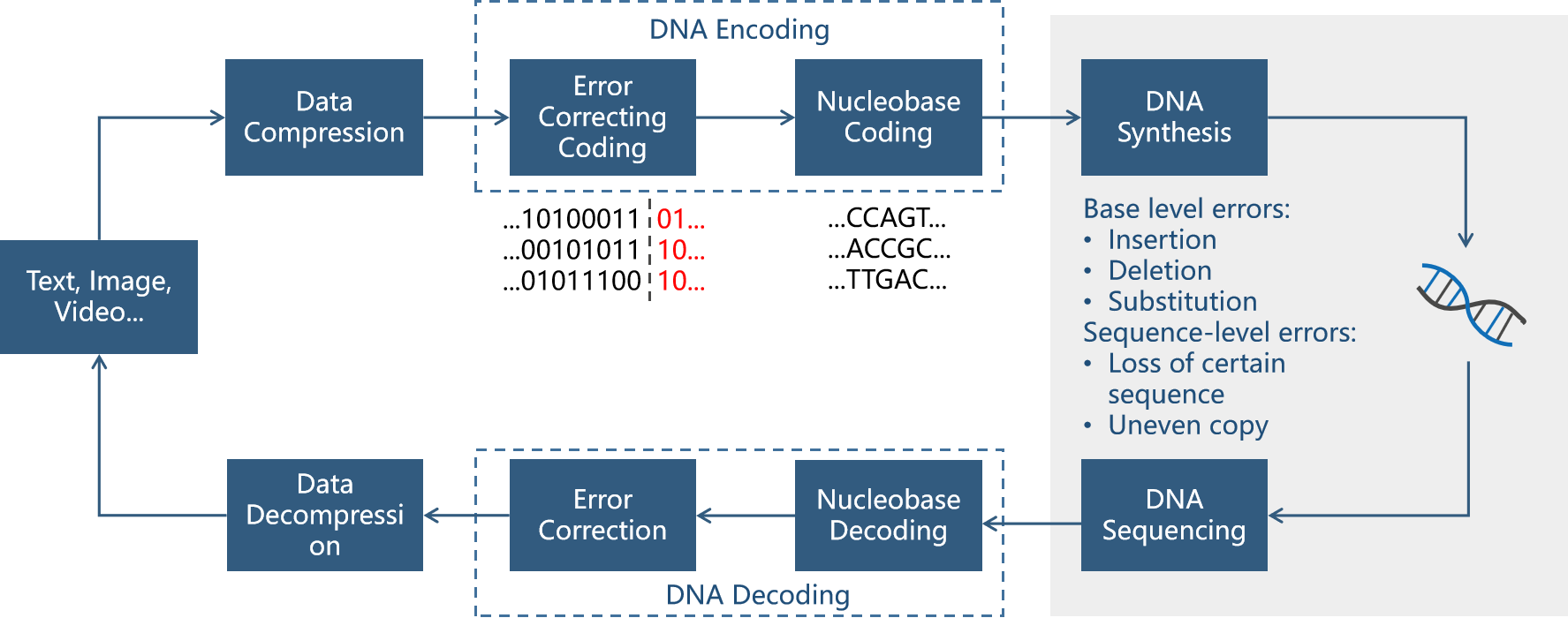}
 	\caption{\textcolor{black}{The main steps of DNA information storage.}}
 	\label{MainStepsOfDNAInformationStorage}
 \end{figure*}


The main processes of DNA information storage are described in Fig. \ref{MainStepsOfDNAInformationStorage}. Specifically, the encoder converts the storage information to a ``$01$” bit sequence; compresses the information into a shorter bit sequence; inserts redundant bits to ensure the error correction performance of DNA information storage; and then converts the encoded information bits into base sequences based on the nucleotide coding rules. This mapping relationship usually has the following types: (1) \textit{binary nucleobase coding}: two nucleobases are employed to represent one information bit, i.e., ``A” and ``C” representing bit ``$1$” while ``T” and ``G” representing bit ``$0$”; (2) \textit{ternary coding}: the information data is first converted into ternary and then converted into bases according to a specific conversion table. By using ternary coding, the selection of the current base depends on the previous base, and the information of the current bit is decoded based on the previous one; (3) \textit{quaternion coding}: the four possibilities of two-bit information are used to represent four bases, which is a coding method with the highest storage efficiency. In addition, some studies use mixed conversion methods \cite{organick2018random}, such as combining binary coding and quaternion coding, etc.

When the DNA sequence is determined, the synthesis of DNA molecules starts manually. The synthesized DNA sequence needs to be inserted with rationally designed primers for subsequent Polymerase Chain Reaction (PCR) process and information retrieval. After that, the DNA will be stored stably for a long time through special schemes, such as solution storage, dry storage, freeze-drying storage with stabilizers and DNA encapsulations \cite{tan2021preservation}.

The DNA molecules are generally not stored optionally, but placed in different pools, and a unique primer sequence is designed in front of each DNA molecule. When it is necessary to extract the information of a certain DNA sequence, the primer of the DNA strand shall be copied by PCR technology, and the target sequence shall be copied by PCR according to the independent pairing and amplification technology of DNA molecules, \textcolor{black}{so as to extract the base sequence based on sequencing technologies mentioned above.}

 \textcolor{black}{After that, the decoder performs the reverse process of the encoder. Specifically, the decoder converts the base sequence into encoded information bits, then performs error correction and decompression, and finally obtains the original data.}

Although DNA storage technology has many advantages, there are still challenges that remain unsolved: (1) \textit{High storage cost}: The huge cost lies in the very high purity of the reagents required for the synthesis; (2) \textit{Read/write latency}: The read/write latency increases with the length of DNA strands. The storage and extraction of information requires DNA synthesis and sequencing. Due to current technical limitations, DNA storage is not suitable for scenes that need to read data immediately \cite{liu2021dna}; (3) \textit{Security requirement}: In order to enhance data reliability and avoid the risk of attacks and thefts, DNA steganography and cryptography are applied to the field of DNA storage \cite{murugan2017cloud}.

\subsection{\textcolor{black}{Errors and Restrictions in DNA storage} }

\label{sec.codingStructure}
How to accurately and efficiently store large amounts of data in DNA is an important issue, and DNA storage differs from traditional information storage, since many errors may occur during the process of DNA synthesis, storage, PCR amplification, and sequencing \cite{heckel2019characterization}. These errors can be divided into two categories: base-level errors and sequence-level errors.

\subsubsection{Base level errors}

\begin{figure*}[!t]
	\centering
    	\includegraphics[width=6in]{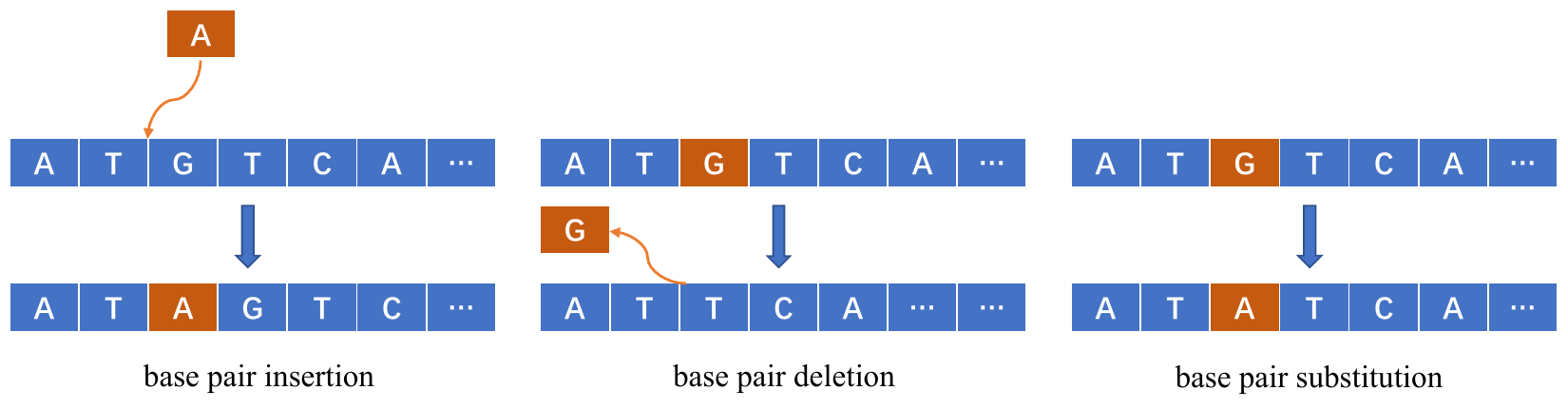}
	\caption{Base-level errors.}
	\label{BaseLevelErrors}
\end{figure*}

During DNA synthesis, there are insertion, deletion and substitution errors in the bases of the DNA sequence, as shown in Fig. \ref{BaseLevelErrors}. Specifically, additional nucleotide molecules are added to the DNA sequence (insertion errors), nucleotide molecules disappear in the expected DNA sequence positions (deletion errors), and the remaining nucleotide molecules replace the original nucleotide molecules at some positions in the DNA sequence (substitution errors) \cite{zhirnov2016nucleic}. 

\begin{figure}[!t]
 	\centering
    \subfigure[Nanopore sequencing]{
    	\includegraphics[width=2in]{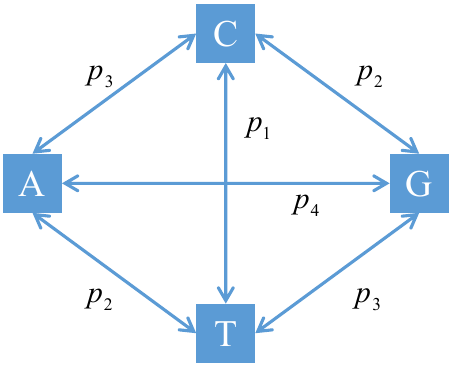}
    	}
    \subfigure[Illumina sequencing]{
    	\includegraphics[width=2in]{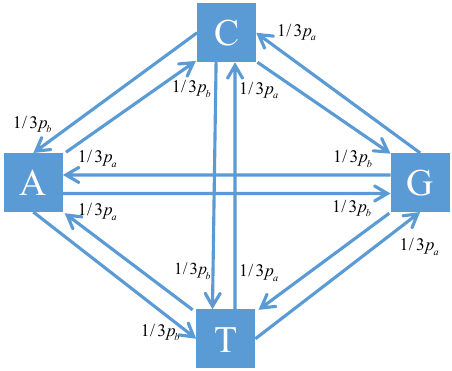}
    	}
    \caption{Substitution error models for nanopore sequencing ($p_{1}$=4$\alpha$, $p_{2}$=$\alpha$, $p_{3}$=0.01, $p_{4}\approx0$, $\alpha$$\in$(0, 0.198)) and Illumina sequencing ($p_{a}$=1.5$\beta$, $p_{b}$=$\beta$, $\beta$$\in$(0, 2/3)).}
    \label{DNASequencerErrors}
\end{figure}

During DNA storage, there is a high probability of hydrolytic damage to the DNA molecule \cite{zhirnov2016nucleic}, as evidenced by the deamination of base C, the conversion of the original G-C to G-U, and the deletion of some nucleotides in the DNA sequence. The sequencing technology directly affects the error rate of the sequencing results. Currently, the mainstream sequencing technologies are mainly second-generation sequencing technologies (high-throughput sequencing, e.g. Illumina sequencing) and third-generation sequencing technologies (nanopore sequencing e.g. Oxford Nanopore Technologies (ONT)). The error rates of the two sequencing technologies vary widely, ranging from 1$\%$-2$\%$ for Illumina sequencing \cite{heckel2019characterization} to  10$\%$-15$\%$ for nanopore sequencing \cite{cretu2017mapping}. The majority of the base-level errors that occur are from the synthesis and sequencing sessions. Fig. \ref{DNASequencerErrors} shows the substitution error models of the nanopore sequencing and the Illumina sequencing \cite{deng2019optimized}. According to the statistical results of reference \cite{press2020hedges}, The proportion of base insertion, deletion and substitution errors in DNA storage were 17$\%$, 40$\%$ and 43$\%$, respectively. 

Error-correcting codes are commonly used to correct substitution errors in DNA sequences, and some of them can also handle insertion and deletion errors between bases. Erasure code has better error correction performance for the problem of molecular inhomogeneity between sequences, and it can also correct the errors within the sequence where only substitution errors exist.

\subsubsection{Sequence level errors}

\begin{figure*}[!t]
	\centering
	\includegraphics[width=5in]{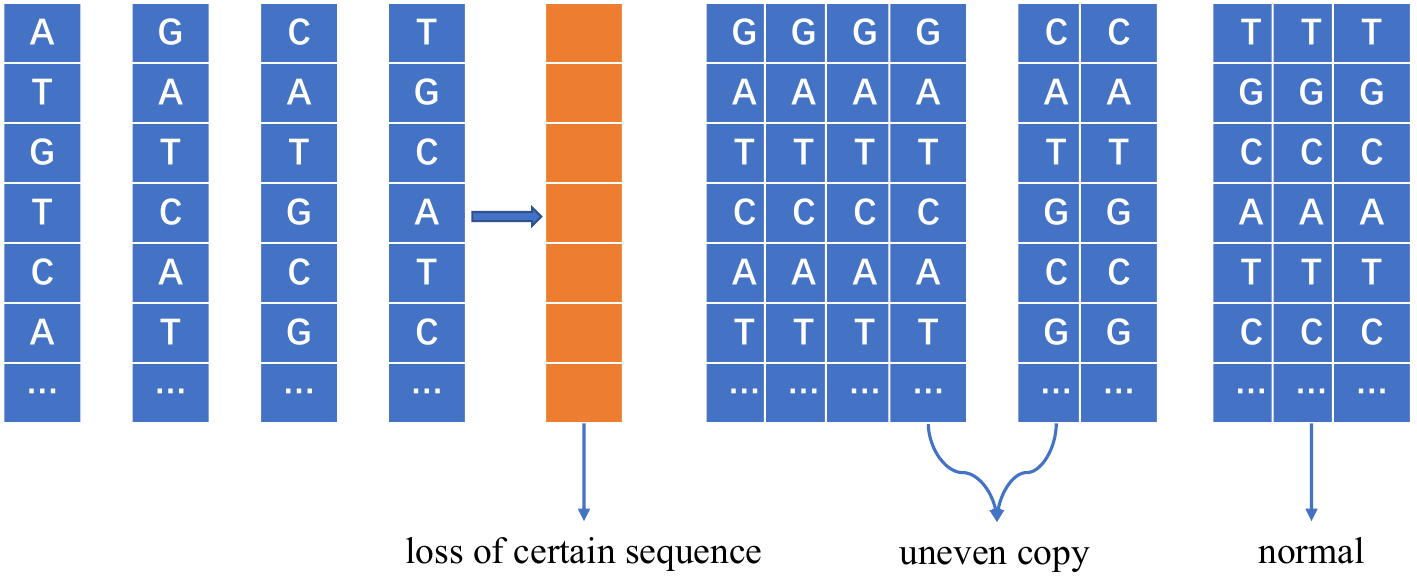}
	\caption{Sequence-level errors.}
	\label{SequenceLevelErrors}
\end{figure*}

Fatal errors can occur not only within the DNA sequence itself but also between sequences. When DNA sequences are synthesized and PCR amplified, sequence uneven copy errors and sequence loss errors may occur. In practice, a DNA sequence has multiple copies in the synthetic pool, and for technical reasons the number of copies of each sequence is different. During subsequent decay, PCR, sampling, and sequencing, this copy number varies widely, so sequence loss and uneven copy occur, as shown  in Fig. \ref{SequenceLevelErrors}. Uneven copy means that some DNA sequences have more copies and some DNA sequences have fewer copies. Finally, the sequence information is recovered through the voting results of multiple sequencing reads.

Erlich proposed the DNA fountain code algorithm to deal with the problem of sequence loss \cite{erlich2017dna}. This method stored data with high DNA storage density at the cost of high encoding and decoding complexity. For error-correcting codes, the solution to the loss of sequence is to convert it into substitution errors. In the encoding process, the binary data in the data block is arranged in columns to form a matrix, and error correction codes are used to add redundant bits to each row of the matrix. During decoding, the binary data in the data block are first arranged according to their respective index values, the missing sequences are replaced with equal lengths of random binary data, and then the bit corresponding to the missing sequences in each row of the matrix is corrected in turn with the error correction code until the missing sequences are fully recovered.

In addition, due to the biological properties of DNA molecules, the encoded DNA sequence should meet the following conditions as much as possible. (1) Avoid the presence of base homopolymers, that is, avoid consecutive arrangements of identical bases. (2) The content of GC bases in a DNA sequence should converge to 50$\%$ as far as possible. (3) The DNA sequence should not have a hairpin structure. These issues can increase the error rate when performing biochemical reactions. In more detail, homopolymers make the DNA sequence more prone to base substitution and deletion, the imbalance of CG content leads to high dropout rates and PCR errors, and the hairpin structure also reduces the efficiency of PCR amplification. Reference \cite{heckel2019characterization} further describes the characteristics of DNA storage channels and the mechanisms by which errors occur.

These issues need to be considered in the base coding stage after channel coding. The base coding strategy in \cite{lochel2022fractal} is quite unique. 
It verifies whether the DNA sequence after base coding conforms to the biological characteristics of DNA, discards the non-conforming sequence, and stores the conforming sequence. This scheme works well for DNA fountain codes. The base coding used in \cite{goldman2013towards} is to convert the ternary sequence into a quaternary base sequence, and the generation of each base is based on the previous base. The production of homopolymers can be avoided.
In \cite{zan2022hierarchical}, the text information and short base sequences are directly mapped one by one, and the selection of each short base sequence takes into account the homopolymer and GC content.
The choice of base code is mainly to reduce the influence of DNA biological characteristics, leading to a decrease in error rates, and to increase the amount of information stored in the base sequence. Of course, each base code should take into account the characteristics of the channel coding scheme used. In the following sections, we will give details on the error correction principle and error correction performance of erasure and error correction codes. 
The base encoding strategy used in the simulations of this paper is quaternary encoding, that is, every two bits are converted into one base.

The main differences between DNA-based molecular storage and traditional communication or storage are as follows: (1) DNA storage channel contains insertion and deletion errors \cite{heckel2019characterization}. (2) A DNA sequence contains multiple copies \cite{yuan2022desp}, which can be used when designing coding methods \cite{xie2023study,jeong2021cooperative}. (3) DNA storage coding needs to meet additional biochemical constraints \cite{ross2013characterizing}. (4) Due to the limitation of synthesis technology, the length of the DNA is limited \cite{hao2020current}, and large-capacity data needs to be split when stored in DNA. Moreover, DNA sequences are mixed in the DNA pool rather than arranged in an orderly manner \cite{heckel2019characterization}, requiring primers and additional coding redundancy to distinguish the sequence of data. (5) Traditional communication often uses soft-decision decoding, while the result of DNA sequencing is definite nucleotides, that is, hard information. Computation of soft information in DNA storage needs to consider channel models and multiple sequencing reads.

\section{Erasure Codes for Nucleotide Deletion and Insertion}


Erasure codes, such as fountain codes, watermark codes and RS codes, are commonly used in the coding process of DNA storage. Specifically, fountain codes prevent the loss of the DNA sequence, while RS  and watermark codes deal with insertion and deletion within sequences. 
In this section, we introduce two typical erasure codes, i.e., fountain codes and watermark codes, and their applications in DNA storage.

\subsection{Fountain Code}

In 2017, Erlich \textit{et al.} \cite{erlich2017dna} first proposed the application of fountain codes in DNA storage, and many studies have further explored DNA fountain codes since then \cite{koch2020dna,jeong2021cooperative,he2023basis}. The benefit of applying fountain codes in DNA storage is that the complex nucleobase coding method that enables DNA oligos to meet biochemical constraints and error-correcting codes for insertion and deletion errors can be ignored.

 In DNA storage, we store the information by synthesizing the DNA oligos. 
 A DNA sequence that conforms to the specified biochemical constraints of the DNA storage (GC content or homopolymer length) is considered valid and can be transmitted; otherwise, it is invalid and discarded. In the process of PCR and sequencing, the valid sequences are exposed to a low dropout rate. A DNA sequence with insertion and deletion errors is easily detected and can be seen as a kind of “dropout”. Accordingly, such channels can be regarded as a cascade connection between a constrained channel and an erasure channel, where fountain codes will work well.

In Erlich’s strategy \cite{erlich2017dna},  the file is first compressed and preprocessed into a binary stream. The fountain code is employed to encode the binary stream into a specific number of segments, referred to as droplets. The $\left\{00,01,10,11\right\}$ in each droplet is mapped to $\left\{\rm{A,T,C,G}\right\}$, respectively. The encoder
screens and discards invalid droplets after converting binary droplets to DNA sequences. Since the droplets are independent of each other and potentially unlimited in number, the loss of some droplets in screening or some error oligos in decoding will not affect the decoding accuracy. Fountain code confers robustness against dropouts. In practice, error correction codes (RS codes) can also be added to each droplet to improve robustness. The decoder can recover the original file as long as it receives enough oligos.

We now give a brief description of the coding and decoding of the Luby transform (LT) \cite{luby2002lt}, which is a type of fountain code. For LT code, each encoding symbols \{$y_1$, $y_2$, ...\} are generated by the source symbols \{$x_1$, $x_2$,..., $x_K$\} through the following steps:

1. Randomly choose the degree $d\in[1, K]$ of the encoding symbol from a degree distribution, such as Robust Soliton Distribution (RSD) \cite{luby2002lt};

2. Uniformly and randomly select $d$ symbols from the source symbols \{$x_1$,$x_2$,\ldots,$x_K$\} as neighbors of the encoding symbol $x_i,i\in[1,K]$;

3. Perform XOR on $d$ neighbors of the encoding symbol to generate one encoding symbol. 

\begin{figure*}[!t]
	\centering
	\includegraphics[width=5in]{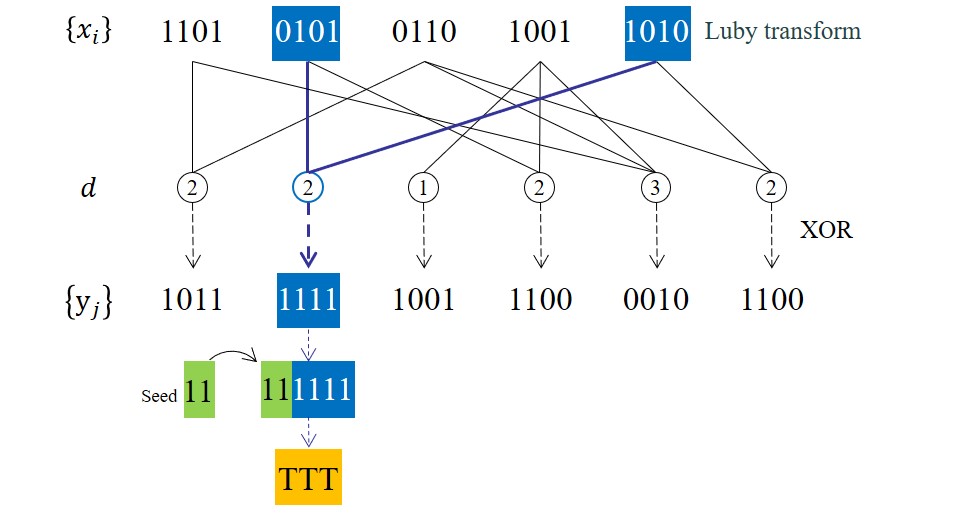}
	\caption{An example of Luby transform codes.}
	\label{AnExampleOfLTCode}
\end{figure*}

A toy example of LT encoding process is shown in Fig. \ref{AnExampleOfLTCode}. Firstly, set $d=2$, uniformly select $x_1=[1101]$ and $x_3=[0110]$, and give the first encoding symbol $y_{1}=x_{1} \oplus x_{3}=[1011]$. The neighbor index set of $y_1$ is $[1,3]$. Then, we calculate other encoding symbols in the same way. Finally, we add seeds in front of the corresponding encoding symbols to get droplets. For example, if we define the seed of $y_2$ as $[11]$, the droplet becomes $[111111]$, and the oligo becomes $[\rm{TTT}]$. 

A typical decoding algorithm for LT codes is the iterative belief propagation (BP) algorithm. During each iteration, the BP algorithm performs the following steps:

1. Find an encoding symbol $y_j$ that has a degree of 1. If there is no such encoding symbol, the decoding fails;

2. Let $x_i=y_j$;

3. Modular addition $x_i$ to all neighbor encoding symbols and replace those encoding symbols with the result. At the same time, the degree of the corresponding encoding symbols should be subtracted by $1$ and the list of neighbors deletes $x_i$.

When decoding the LT coded sequence exemplified in Fig. \ref{AnExampleOfLTCode},  we first find the encoding symbol $y_3$ that is only connected to $x_4$. Secondly, we have $x_4=y_3$. Thirdly, modular addition is performed on $x_4$ to its neighbors $y_4$ and $y_5$. Update the corresponding degree. Finally, we delete the corresponding lines. 
Repeating the above steps, we recover ${\hat{x}_{2}}=y_{4} \oplus {x}_{4}=[1100] \oplus[1001]=[0101]$ in the 1st iteration,
$ {\hat{x}_{5}}=y_{2} \oplus {x}_{2}=[1111] \oplus[0101]=[1010]$ in the 2nd iteration,
${\hat{x}_{3}}=y_{6} \oplus {x}_{5}=[1100] \oplus[1010]=[0110]$ in the 3rd iteration, and 
$ {\hat{x}_{1}}={y}_{1} \oplus {x}_{3}=[1011] \oplus[0110]=[1101]$ in the $4$-th iteration.

If the original data consists of $K$ source symbols, then each encoding symbol can be generated on average by $O(\rm{ln}(K/\delta))$ symbol operations and the $K$ original source symbols can be recovered from any $K+O(\sqrt{K}\rm{ln}^2(K/\delta))$ of the encoding symbols with probability $1-\delta$ by on average $O(K\cdot \rm{ln}(K/\delta))$ symbol operations, where a symbol operation is either an exclusive-or of one symbol into another or a copy of one symbol to another \cite{luby2002lt}.


\subsection{Watermark Code}

Davey and Mackey \cite{davey2001reliable} proposed a class of codes, cascaded by LDPC codes and watermark codes, to correct insertion, deletion, and substitution errors in the insertion-deletion-substitution (IDS) channel, which is called Davey-MacKay (DM) construction.
The DM construction can be used for DNA coding \cite{chen2021artificial,maarouf2022concatenated}, and its inner code is the watermark code, which is can be used to find errors that happen in a certain order in the IDS channel.

The core idea of the watermark code is to XOR the information sequence and the watermark sequence, a pseudo-random sequence, before transmission and then compare the accepted sequence with the known watermark sequence at the receiving end to obtain the actual soft information of the information sequence using a hidden Markov model (HMM) forward-backward algorithm \cite{davey2001reliable}. To better handle insertion-deletion errors, define the bit drift $x_{i}$ (the hidden state in the HMM \cite{rabiner1986introduction}) as
\begin{equation}
	\label{}
	x_{i}=n_{\text{ins}}^i-n_{\text{del}}^i,
\end{equation}
where $n_\text{ins}^i$ and $n_\text{del}^i$ represent the number of inserted and deleted bits before position $i$, respectively.

\begin{figure}[!t]
	\centering
	\includegraphics[width=0.48\textwidth]{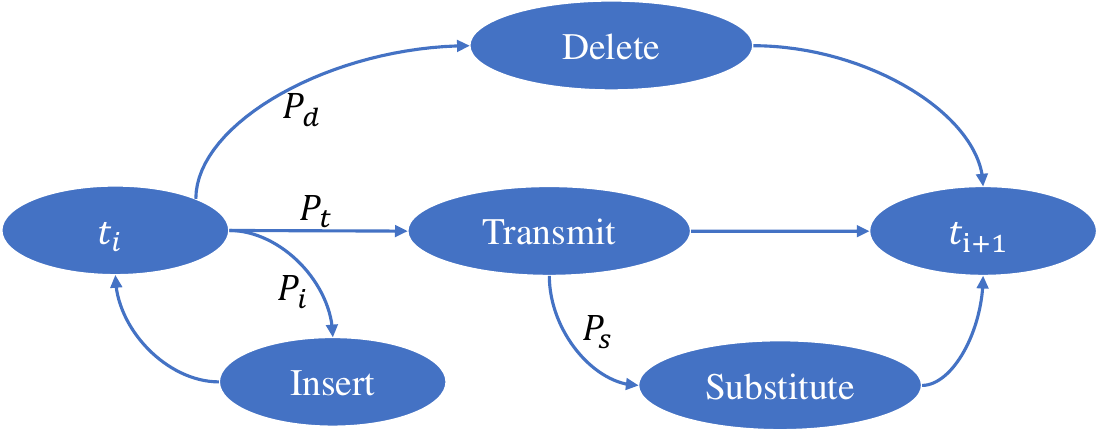}
	\caption{The IDS channel model.}
	\label{ErrorsDuringTheConversion}
\end{figure}

The sequence $r$ received by the decoder is the observed sequence in the HMM. \textcolor{black}{When receiving the $i$-th bit from position $i$ ($t_{i}$) to position $i+1$ ($t_{i+1}$), we define the probabilities of insertion, deletion, transmission and substitution as $P_{i}$, $P_{d}$, $P_{t}$ and $P_{s}$ respectively, as shown in Fig. \ref{ErrorsDuringTheConversion}. It is worth noting that the probabilities of inserting bits are independent.}

For watermark sequences, the information sequence is a noise. In order to avoid the effect of noise when the information sequence is superimposed on the watermark sequence, the information sequence needs to be sparse before XOR processing.
The information sequence to be encoded in the watermark code refers to the sequence encoded by the outer code in the DM construction. Hence, sparse processing is to perform sparse processing on the codeword $d$ of the outer code, mapping each $q$-ary symbol of $d$ to $n$ bits with lower density.
The sparsity $f$ is defined as the average density of the sequence after sparsing. When performing sparsification, it is necessary to reduce $f$ as much as possible. At the decoder, in addition to the interference caused by the superposition of information sequences, the channel also contains substitution error noise, so the effective substitution error probability $P_{f}$ is given as
\begin{equation}
	\label{}
	P_{f}=f\left(1-P_{s}\right)+(1-f) P_{s}.
\end{equation}

Additionally, the transition probability $P_{a b}=P(x_{i+1}=a|x_{i}=b)$ is defined as the likelihood of the transition between two adjacent positions with different offsets, which can be expressed as
\begin{equation}
	\label{WatermarkTransitionProbability}
	\begin{array}{lr}
		P_{a b}=\left\{\begin{array}{l}
			P_{d}, b=a-1; \\
			\alpha_{I} P_{i} P_{d}+P_{t}, b=a; \\
			\alpha_{I}\left(\left(P_{i}\right)^{b-a+1} P_{d}+\left(P_{i}\right)^{b-a} P_t\right),  \\ 
            \qquad a<b<a+I;  \\
			\alpha_{I}\left(P_{i}\right)^{I} P_{t}, b=a+I; \\
			0, \text{ otherwise, }
		\end{array}\right. 
	\end{array}
\end{equation}
where $a$ is the drift at $t_{i}$, $b$ is the drift at $t_{i+1}$, $I$ represents the drift limit of adjacent bits,
and $\alpha_{I}$ is defined as
\begin{equation}
	\label{}
    \alpha_{I}=1 /\left[1-\left(P_{i}\right)^{I}\right],
\end{equation}
where $P_{a b}$ represents two situations: (1) $b-a+1$ bits are inserted between $t_{i}$ and $t_{i+1}$, and the bit of $t_{i+1}$ is deleted; (2) $b-a$ bits are inserted between $t_{i}$ and $t_{i+1}$, and the bit of $t_{i+1}$ isn't deleted. Hence, assuming that the probability of inserting $b-a+1$ bits is $\lambda$, and the probability of inserting $b-a$ is $\mu$, $P_{a b}$ can also be expressed as
\begin{equation}
	\label{SimpleformofPab}
    P_{ab}=\lambda P_{\mathrm{d}}+\mu P_{\mathrm{t}}.
\end{equation}

The forward-backward algorithm of the Markov model also computes the emission probability $Q_{a b}^{i}(s)$ from $x_i=a$ to $x_{i+1}=b$, which can be expressed as
\begin{equation}
Q_{a b}^i(s)=\left\{\begin{array}{cl}
& \frac{\lambda P_{d} / 2^{b-a+1}+\mu P_{t}\left(1-P_{f}\right) / 2^{b-a}}{P_{a b}} , 
\\ & \qquad s^*=w_i;  \\
& \frac{\lambda P_{d} / 2^{b-a+1}+\mu P_{t} P_{f} / 2^{b-a}}{P_{a b}} , 
\\ & \qquad s^*=w_i \oplus 1 ,
\end{array}\right.
\end{equation}
where $s$ is the output string from $x_i=a$ to $x_{i+1}=b$, $s^{*}=s_{b-a+1}$ is the received bit associated with the watermark bit $w_{i}$.

The forward probability $F_{i}(y)$ represents the probability of that $x_i$ is $y$ and the first $i-1+y$ bits are received. The recursive expression for the forward probability from front to back is expressed as
\begin{equation}
\begin{aligned}
	\label{ForwardProbability}
F_{i}(y)&=P\left(u_{1}, \ldots, u_{i-1+y}, x_{i}=y \mid \mathcal{H}\right)\\
&=\sum_{a=y-I}^{y+1} F_{i-1}(a) P_{a y} Q_{a y}^{i-1}\left(u_{i-1+a}, \ldots, u_{i-1+y}\right).
\end{aligned}
\end{equation}

The calculation of backward probability $B_{i}(y)$ is similar to the forward probability in (\ref{ForwardProbability}), which can be expressed as
\begin{equation}
\begin{aligned}
	\label{BackwardProbability}
B_{i}(y)&=P\left(u_{i+y}, \ldots \mid x_{i}=y, \mathcal{H}\right)\\
&=\sum_{b=y-I}^{y+1} B_{i+1}(b) P_{y b} Q_{y b}^{i}\left(u_{i+y}, \ldots, u_{i+b}\right).
\end{aligned}
\end{equation}

The watermarking algorithm provides the final output of the likelihood function $P\left(r \mid d_{i}, \mathcal{H}\right)$ of the outer code's codeword $d$ for further processing by using the forward-backward algorithm of the HMM described in the previous section, which can be expressed as
\begin{small}
\begin{equation}
\begin{aligned}
	\label{watermark likelihood function}
&P\left(u \mid d_{i}, \mathcal{H}\right)=\sum_{x_{i_{-}}, x_{i_{+}}} F_{i_{-}}\left(x_{i_{-}}\right) B_{i_{+}}\left(x_{i_{+}}\right) \cdot \\
& \qquad P\left(u_{(i_{-}+x_{i_{-}})}, \ldots, u_{(i_{+}+x_{i_{+}-1})}, x_{i_{+}} \mid x_{i_{-}}, d_{i}, \mathcal{H}\right) ,
\end{aligned}
\end{equation}
\end{small}

where $\mathcal{H}$ contains the $P_{i}$, $P_{d}$, $P_{t}$ and  $P_{s}$ in the channel, the watermark sequence information $w$, the effective replacement probability $P_{f}$ and the average density $f$ of the information sequence. $i_{-}=n \times i$,  $i_{+}=n \times (i+1)$. $d_i$ is sparse into $n$ bits. $P\left(u \mid d_{i}, \mathcal{H}\right)$ represents the probability that the received bits corresponding to $d_i$ are $u_{(i_{-}+x_{i_{-}})}, \ldots, u_{(i_{+}+x_{i_{+}-1})}$. $F_{i_{-}}$ and $B_{i_{+}}$ represent the probability of positioning $d_i$ as $u_{(i_{-}+x_{i_{-}})}, \ldots, u_{(i_{+}+x_{i_{+}-1})}$. \ $P\left(u_{(i_{-}+x_{i_{-}})}, \ldots, u_{(i_{+}+x_{i_{+}-1})}, x_{i_{+}} \mid x_{i_{-}}, d_{i}, \mathcal{H}\right)$ represents the codeword $d_i$ information and the calculation method is the forward calculation from $x_{i_{-}}$ to $x_{i_{+}}$ in (\ref{ForwardProbability}), where $w_{i}$ is replaced with  $w_{i} \oplus 1$ and  $P_{f}$ with $P_{s}$.

\begin{table*}[]
    \centering
    \caption{Definitions of Notations}
    \begin{tabular}{cl}
       \toprule
       Notation & Definition \\
       \hline
       $k$ & information length before LT or watermark encoder  \\
       $\epsilon$ & the ratio of the number of source symbols to the number of encoding symbols in the LT code  \\
       $N$ & block length  \\
       $m$ & The symbols in RS are the elements in the Galois field $GF(2^m)$.  \\
       $r$ & code rate  \\
       $\alpha$ & the probability of base substitutions in nanopore sequencing model  \\
       $\beta$ & the probability of base substitutions in Illumina sequencing model  \\
       FER & frame error rate  \\
       BER & bit error rate  \\
       $u$ & the received sequence before watermark decoder \\
       $P_{l}$ & the loss probability of DNA oligos  \\
       $P_{i}$ & the probability of base insertions  \\
       $P_{d}$ & the probability of base deletions  \\
       $P_{s}$ & the probability of base substitutions  \\
       $P$ & the error probability consists of several items from $P_{l}$, $P_{i}$, $P_{d}$ and $P_{s}$ \\
       \bottomrule
    \end{tabular}
    \label{Definitions of Notations}
\end{table*}

\begin{figure}[!t]
 	\centering
    \subfigure[]{
    	\includegraphics[width=3in]{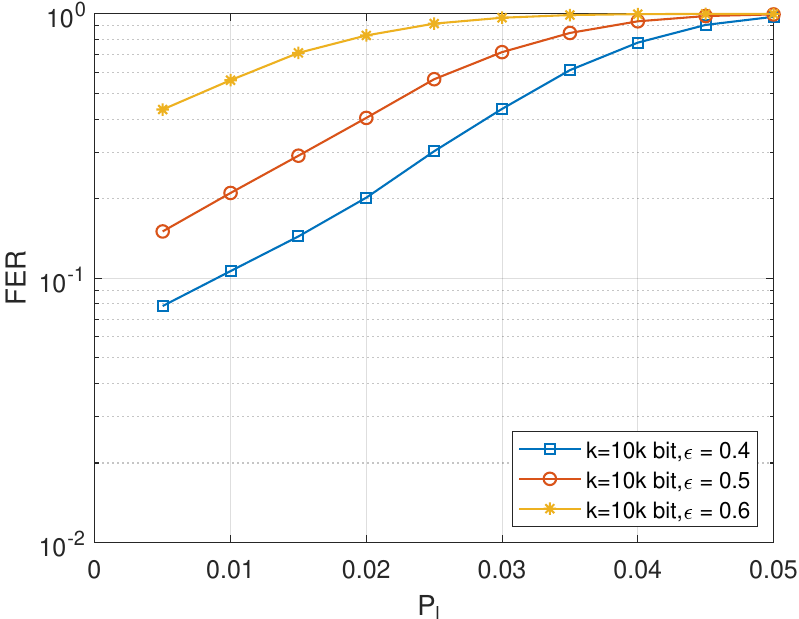}
    	\label{LT_Simulation}
    	}
    \subfigure[]{
    	\includegraphics[width=3in]{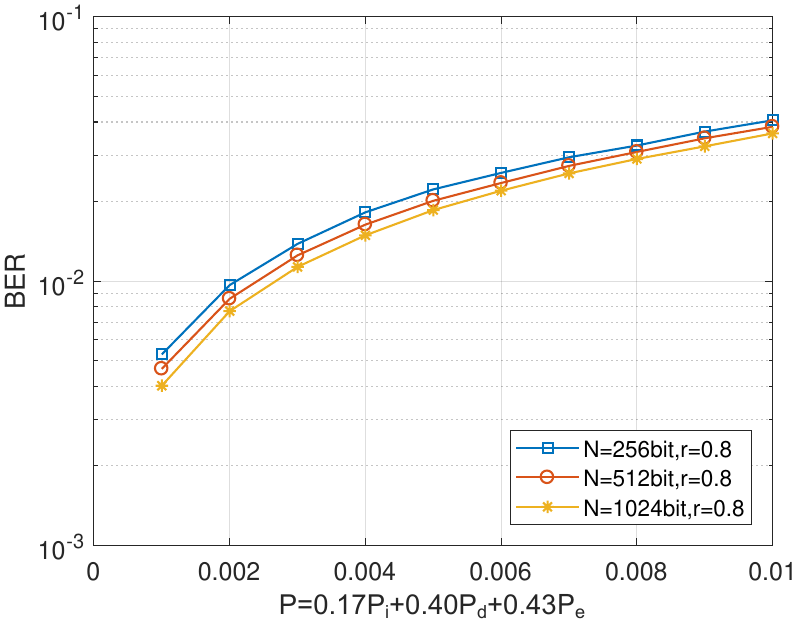}
    	\label{Watermark_Simulation}
    	}
    \caption{(a) FER performance of LT codes with sequence loss; (b) BER performance of watermark codes with insertion, deletion, and substitution errors.}
    \label{Watermark_LT_Simulation}
\end{figure}

To reduce the complexity, we limit the maximum drift to $M$. Assume that the watermark code rate is $r=p/t$ and the complexity of calculating forward or backward probability once is $O(1)$. The complexity of the watermark decoder is $O(2MN+p2^{t}M^{2}(N/t))$, where $MN$ represents the complexity of calculating the total of the forward or backward probability, and $O(p2^{t}M^{2}(N/t))$ represents the complexity of calculating the final output of the likelihood function $P\left(r \mid d_{i}, \mathcal{H}\right)$ \cite{zhang2017iterative}.

We also provide some numerical results when employing LT or watermark codes in DNA storage. Table \ref{Definitions of Notations} lists the definitions of relevant notations in subsequent simulation diagrams. For LT codes and LT-RS codes, the frame error rate (FER) in the simulation diagram represents the success probability of a single decoding, that is, a single simulation. Fig. \ref{LT_Simulation} shows the FER performance of LT codes over the erasure channel. Fountain codes handle sequence loss errors better and recover more oligos when  $\epsilon$ is lower and $k$ is larger. When the information length is very large ($K$ reaches tens of thousands and $k$ reaches millions of bits), a high decoding success rate can be obtained with small redundancy. In the fountain code, obtaining the information of a specified source symbol usually requires waiting until the decoding is completely successful, and successful decoding requires the use of more than $K$ encoding symbols. This means that it is not convenient to achieve random access to a small piece of data in DNA when using fountain codes. Fig. \ref{Watermark_Simulation} shows the bit error rate (BER) of watermark codes when handling insertion and deletion errors. The insertion, deletion, and substitution probabilities account for 17$\%$, 40$\%$ and 43$\%$ of the total error probability, respectively. As shown in Fig. \ref{Watermark_Simulation}, the length of the watermark code does not affect the performance of the watermark code significantly. Watermark codes correct insertion and deletion errors, but there are still substitution errors in the output of the watermark code. 

\section{Error Correction Codes for Nucleotide Substitution}
In section \uppercase\expandafter{\romannumeral3}, we have discussed how to employ erasure codes to correct insertion and deletion errors. However, they cannot deal with the substitution errors. Therefore, in this section, we introduce the ways of employing error correction codes to correct substitution errors in DNA storage. Three typically employed codes, i.e. BCH codes, LDPC codes and polar codes, are introduced in Sections \uppercase\expandafter{\romannumeral4}-A to \uppercase\expandafter{\romannumeral4}-C, respectively.


\subsection{BCH Code}

The BCH code is an important class of cyclic codes that can correct multiple random errors. It has strong applicability and flexible parameter selection with a wide selection of code rates $R=k/n$ and code lengths up to hundreds of bits. The most common one is the binary BCH code. However, due to its relatively smaller code rate, it is generally only used to protect some important segments of oligos \cite{blawat2016forward}. A typical class of BCH codes for DNA storage is the RS code. RS codes can be used alone to protect entire oligos \cite{grass2015robust} or in combination with other encoding methods such as fountain codes \cite{erlich2017dna}.

\subsubsection{Binary BCH code}

Binary BCH codes are one of the simplest and most common linear block codes. The main parameters of binary BCH codes include the code length $n$, the number of information symbols $k$, error-correcting capability $t$ and minimum Hamming distance $d_0$ where we have $n=2^m-1$, $k\geq n-mt$, $t<(2^m-1)/2$, $d_0\geq2t+1$. 

The encoding process of binary BCH code is detailed below:

The generator polynomial of the BCH code can be expressed as
\begin{equation}
	\label{}
g(x)=\operatorname{LCM}\left[m_{1}(x), m_{3}(x), \ldots, m_{2 t-1}(x)\right],
\end{equation}
where $\rm{LCM[\cdot]}$ represents the least common multiple (LCM) and $m_{i}(x)$, $i=1,3,...,2t-1$ is the minimal polynomial that can usually be obtained directly from the minimal polynomial table. The code polynomial can be obtained by the following steps: (1) Divide the polynomial $x^{n-k}m(x)$ by the generator polynomial $g(x)$ to get the remainder check polynomial $r(x)$; (2) Add $r(x)$ to $x^{n-k}m(x)$ to get the encoded BCH code $c(x)$.

Take the BCH $(15,7)$ code as an example. It can correct $2$ errors, and its generator polynomial is $g(x)=(x^4+x+1)(x^4+x^3+x^2+x+1)$. Assuming that the binary sequence to be sent is $[1010101]$, which can be represented as an information polynomial, $m(x)=x^6+x^4+x^2+1$. The remainder check polynomial is $r(x)=x^8\times m(x){/g(x)=x}^7+x^6+x^5+x^2+1$ and the codeword $C$ is $[101010111100101]$.

The corresponding decoding process of binary BCH codes includes four steps : (1) Calculate the syndromes ${S}_{{j}}=\sum_{{i}=0}^{{n}-1} {R}_{\mathrm{i}} \alpha^{{ij}}, {j}=1,2, \ldots, 2{t}$ with the received word. Due to the transmission errors, not all syndromes remain zero. (2) Define the error location polynomial as $\sigma(x)=\left(1-x_{1} x\right)\left(1-x_{2} x\right) \cdots\left(1-x_{i} x\right)=1+\sum \sigma_{i} x^{i}$, where $x_i$ corresponds to the $i$th error location and $\sigma(x)$ is commonly calculated by the Berlekamp-Massey (BM)algorithm or Euclid algorithm. (3) Calculate the root of $\sigma(x)=0$, which is usually obtained by the Chien search algorithm, and then invert the root to get the error position. (4) Correct the errors by inverting the binary bits in the position where errors occur.

For example, suppose the received codeword $R$ is $[101010011100111]$. The syndromes are  $S_{j}=1+\alpha^{j}\alpha^{2 J}+\alpha^{5 j}+\alpha^{6 j}+\alpha^{7 j}+\alpha^{10 j}+\alpha^{12 j}+\alpha^{14 j}$ , that is, $S_{1}=\alpha^{10}, S_{2}=\alpha^{5}, S_{3}=\alpha, S_{4}=\alpha^{10}$. The corresponding error position polynomial is $\sigma(x)=1+\alpha^{10}x+\alpha^9x^2$, and its roots are $x_1^{-1}=\alpha^7$ and $x=\alpha^{14}$. Therefore, the error positions are $x_1=\alpha^8$ and $x_2=\alpha^1$, and the recovered codeword $\hat{C}$ is $[101010111100101]$.
 
\subsubsection{Reed-Solomon code} 
 
The RS code is a special non-binary BCH code, and its operation is carried out in Galois Field $GF(q=p^m)$. RS code in the $GF(2^m)$ domain is most commonly used in DNA storage. The main difference between RS codes and binary BCH codes is that BCH codes are encoded and decoded in units of bits, while RS codes are encoded and decoded in units of symbols. Therefore, BCH codes can correct random errors, and RS codes can correct a combination of random and burst errors. A q-ary RS code with a code length of $n=q-1$ symbols (that is, $(q-1)m$ bits) and a redundancy of $2t$ symbols can correct $t$ errors.

The generator polynomial of RS($n,k$) codes can be expressed as
\begin{equation}
\begin{aligned}
	\label{}
g(x)&=(x-\alpha)\left(x-\alpha^{2}\right) \cdots\left(x-\alpha^{2 t}\right)\\
&=\prod_{i=1}^{2 t}\left(x-\alpha^{i}\right),
\end{aligned}
\end{equation} 
where $\alpha^i$ is the element in the Galois field $GF(2^m)={0,\alpha^0,\alpha^1,...,\alpha^{2m-2}}$ . After obtaining $g(x)$, the codeword can be calculated by using the encoding method of BCH codes that we mentioned earlier.

Consider the RS(7, 3) code over $GF(2^3)$, which can correct $2$ symbol ($6$ bits) errors. The primitive polynomial is $p(x)=x^3+x+1$, and the generator polynomial is $g(x)=(x-\alpha)(x-\alpha^2)(x-\alpha^3)(x-\alpha^4){=x}^4+\alpha^3x^3+x^2+\alpha x+\alpha^3$. Suppose the binary message to be sent is $[011 010 010]$, which can be converted to elements $(\alpha^3,\alpha,\alpha)$ in $GF(2^3)$. Hence, the codeword is $(\alpha^3,\alpha,\alpha,\alpha^2,\alpha^3,\alpha^2,\alpha^6)$, which can be represented as a binary sequence $[011 010 010 100 011 100 101]$.

\begin{figure*}[!t]
	\centering
	\includegraphics[width=6in]{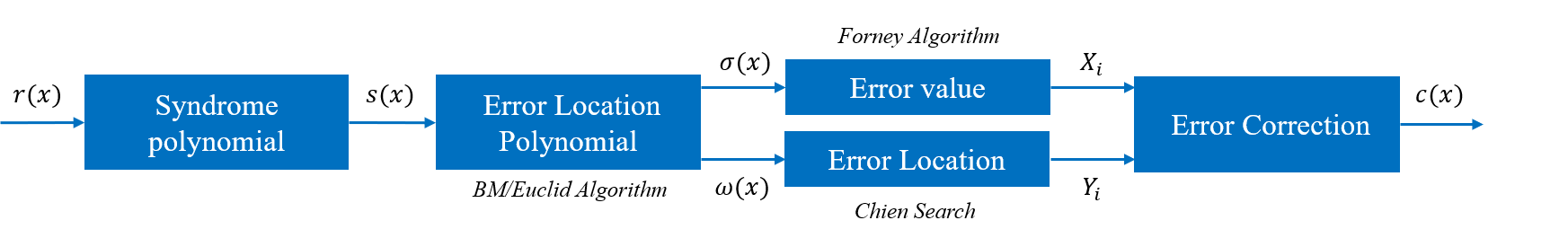}
	\caption{General decoding steps of RS code.}
	\label{GeneralDecodingStepsOfRSCode}
\end{figure*}

The decoding process of RS codes is shown in Fig. \ref{GeneralDecodingStepsOfRSCode}. The complexity of the BM algorithm, Chien search algorithm, and Forney algorithm are $O(n^2)$, $O(nt)$ and $O(t^2)$ respectively \cite{garrammone2013decoding, chen2008complexity}. In the last step, there are only two values for each position of the binary code. However, each position of RS codes has multiple possible values, and the correct value needs to be calculated by Forney's algorithm and the error value polynomial $\omega(x)$. With some modifications to the above decoding method, the RS code can also be used to correct deletion errors. It should be noted that sometimes RS code is only used for error detection and not for error correction. When it is used together with the fountain code, error correction may lead to more errors.
 
RS codes in DNA storage typically appear in three forms: (1) Using a single one-dimensional RS code to protect the payload, such as \cite{organick2018random}, which will bring about 15\% redundancy. In order to strengthen the protection of seeds, \cite{anavy2019data} adds different RS codes to payloads and seeds. (2) Using RS codes as outer codes together with other codes that can correct insertion and deletion errors as inner codes, such as fountain codes \cite{erlich2017dna,anavy2019data} and HEDGES \cite{press2020hedges}. In this case, the RS code will add around 10 nucleobases of redundancy to each oligo. (3) Using the RS code as the inner code and the outer code, respectively \cite{grass2015robust,antkowiak2020low}, which will bring more than $30$\% redundancy.

\begin{figure}[!t]
	\centering
	\includegraphics[width=3in]{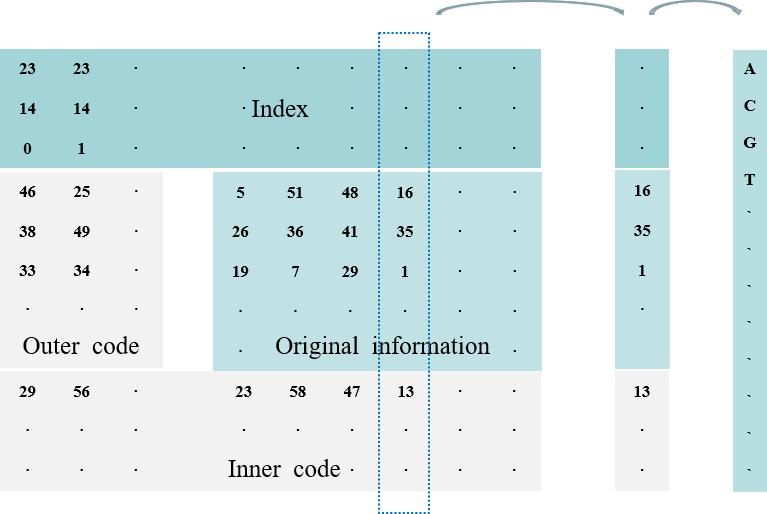}
	\caption{The structure of RS-RS concatenated codes.}
	\label{StructureOfRSConcatenatedCodes}
\end{figure}

The first two methods are common usage of RS codes in traditional coding. In the third case, RS codes are used to encode data in both the horizontal and vertical directions, so that not only errors within the sequence, but also missing errors of the entire sequence can be corrected. As shown in Fig. \ref{StructureOfRSConcatenatedCodes}, the coding steps are: (1) arranging the original information in blocks; (2) performing the outer encoding step, encoding each row of data with RS code and adding redundant bits to the left side of data blocks; (3) adding an index to each column of data; and (4) performing the inner encoding step, encoding each column of data with a second RS code. The basic idea of this method is to convert the loss of the DNA sequence into substitution errors and then correct them through the error correction code. When decoding, we first arrange the received sequences according to their index, then decode the outer code. If there is a missing sequence, replace it with random data of equal length, and then decode the inner code.

\subsubsection{Simulations}

Now we demonstrate some simulation results for RS-coded DNA storage systems for the above three cases, respectively. 

\begin{figure}[!t]
	\centering
	\subfigure[Simulation of RS code with substitution errors]{
	  \includegraphics[width=3in]{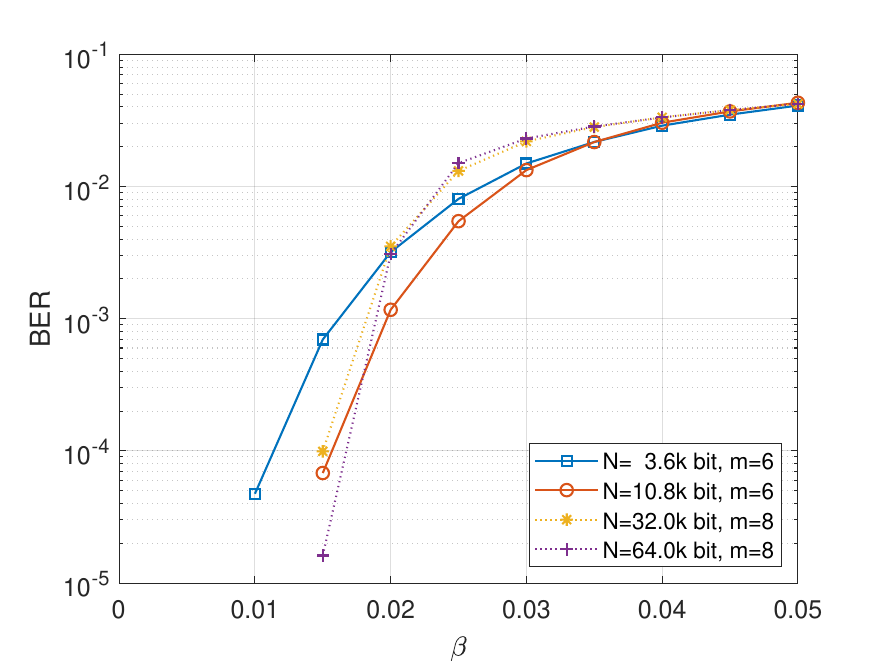}
	  \label{RSwithbeta}}
	\subfigure[Simulation of LT-RS and RS-RS codes with sequence loss and substitution errors]{
	  \includegraphics[width=3in]{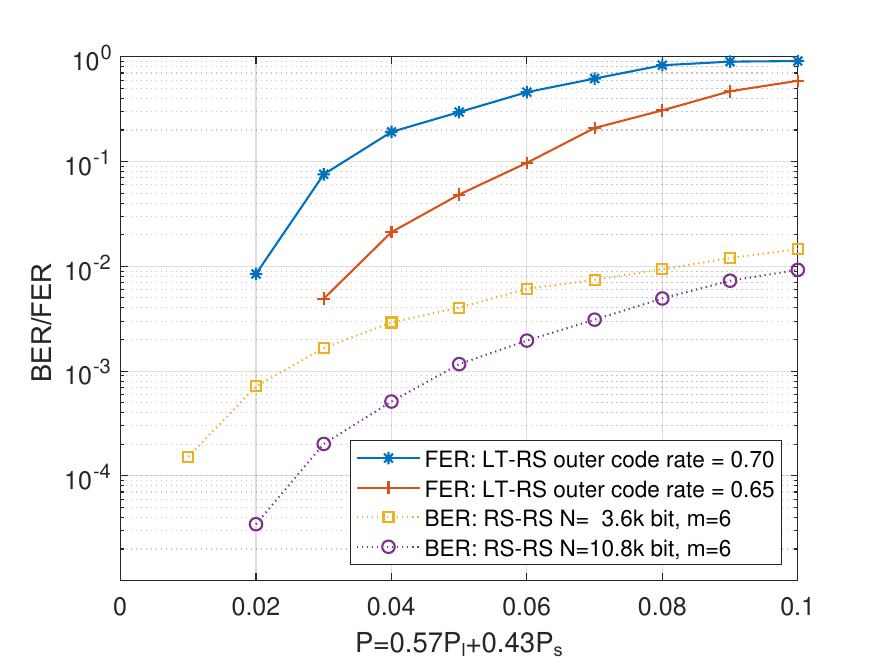}
	\label{RSRSLTWithPePs}}
   \caption{Simulation of RS-coded DNA storage systems.}
   \label{RS_Simulation}
\end{figure}

Initially, an examination of the BER under varying substitution probabilities and code lengths is conducted, as illustrated in Fig. \ref{RSwithbeta}. Here, the coding rate remains constant at $0.68$. The focus is particularly on channels subjected to nucleobase substitutions. A notable observation is that an increase in $\beta$ leads to a surge in the BER. Interestingly, for elevated values of $\beta$, the BER shows marginal fluctuations across different code lengths.

RS codes, when amalgamated with fountain codes, hold the potential for safeguarding oligonucleotides within DNA fountains. Specifically, the LT code operates as the internal code, while the RS code functions as the external code. Fig. \ref{RSRSLTWithPePs} portrays the FER) of LT-RS concatenated codes under distinct error probabilities denoted by $P$. Notably, these errors predominantly comprise substitution errors (57\%) and sequence loss errors (43\%). The stipulated DNA sequence length stands at $156$ nt, accompanied by $\epsilon$ fixed at 0.4 and $k$ benchmarked at 10$k$ bits. The adept fusion of LT and RS codes proves competent in addressing both sequence loss and substitution anomalies. Additionally, Fig. \ref{RSRSLTWithPePs} presents the BER corresponding to RS-RS concatenated codes within an identical channel environment. Comprehensive specifics of both the internal and external codes are detailed in Table \ref{tab:table2}. During these simulations, shortened RS codes are deployed, with messages initially appended with zeros prior to encoding. Post-encoding, extraneous zeros are eliminated to reduce the codeword length. Consistently, each oligonucleotide length is approximated at 150nt, with a singular symbol serving as the index length, and valid data covering a span of nearly 68\%. It's evident that the concatenated model exhibits enhanced resilience against substitution and sequence loss errors, particularly when the block length, $N$, is considerable. Emphasizing data integrity, the LT-RS approach aims for an expansive oligonucleotide recovery, steering clear of generating sequences that contravene biological standards, thereby diminishing error potential. In contrast, RS-RS codes manifest commendable prowess in rectifying base errors.
 
\begin{table*}[!t]
 	\caption{Specification of codes\label{tab:table2}}
 	\centering
 	\begin{tabular}{|c|c|c|}
 		\hline
 		Outer code & Inner code & Block length\\
 		\hline
 		(12,10)RS code over $GF(2^6)$ & (50,42)RS code over $GF(2^6)$ & 3600\\
 		\hline
 		 (36,30)RS code over $GF(2^6)$ & (50,42)RS code over $GF(2^6)$ & 10800\\
 		\hline
 	\end{tabular}
\end{table*}

\subsection{LDPC Code}
 
LDPC code is a kind of linear block code with a sparse check matrix that has low decoding complexity, a flexible structure, and good performance approaching the Shannon limit. LDPC code has
better performance 
at long code lengths than short ones and can be used on long DNA strands such as artificial chromosomes of yeast \cite{chen2021artificial}. It can be used alone \cite{fei2019ldpc,yim2014essential} or in conjunction with other codes that corrects nucleobase insertion and deletion errors, such as the watermark code \cite{chen2021artificial}. 

\begin{figure*}[!t]
	\centering
	\includegraphics[width=5in]{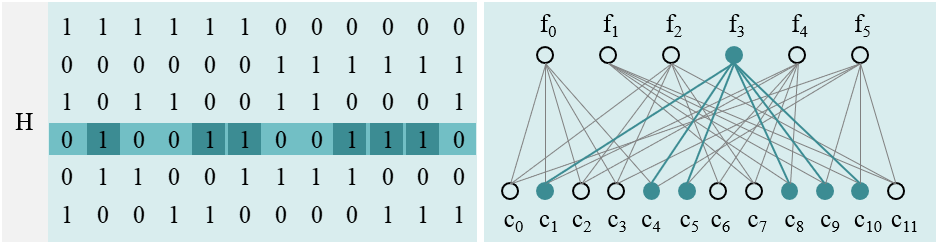}
	\caption{Check matrix and Tanner diagram of LDPC(12,3,6).}
	\label{CheckMatrixAndTannerDiagram}
\end{figure*}

The check matrix $H$ of the LDPC code is a sparse matrix that can be used to uniquely determine the codeword. Only a small part of the element values in $H$ are $1$, and most of the remaining element values are $0$. The row weight in $H$ refers to the number of non-zero elements in the row, and the column weight refers to the number of non-zero elements in the column. If the row weight and column weight of the parity check matrix are fixed, the code is called a regular LDPC code, otherwise, it is called an irregular LDPC code. In addition to a sparse check matrix, LDPC codes are also commonly represented by Tanner graphs. In the Tanner graph, each variable node corresponds to a bit in the codeword, and each check node corresponds to a row of the matrix $H$. A path is defined as a finite sequence of alternating nodes and connections. A cycle is formed when the start node and the end node of a path coincide, such as the path $c_0\rightarrow f_2\rightarrow c_3\rightarrow f_0\rightarrow c_0$ in Fig. \ref{CheckMatrixAndTannerDiagram}. Among all the cycles in the graph, the shortest cycle length is called the girth of the Tanner graph. The girth will affect the convergence of iterative decoding, so the appearance of short cycles should be avoided as much as possible.

There are a variety of methods to construct a parity check matrix $H$, which are mainly divided into two categories: random construction, such as Gallager code, Mackay code, and Davey code, and structured construction, such as finite geometric construction and combinatorial design. Take the Gallager code as an example: a (N, J, K) LDPC code corresponds to the matrix $H(HJ \times HK)$, where $N = HK$, and $J$ and $K$ are row and column weights. First construct $j$ sub-matrices $H_j(H \times  HK),j=1,2,...,J$, then the matrix $H$ can be expressed as: $H=[H_1,H_2,...,H_J]^T$. Each column of sub-matrices $Hj$ has only one non-zero element, and each row has $h$ non-zero elements. The $k$ non-zero elements of each row of $H_1$ are arranged as follows: the position of the non-zero elements of the $i$-th row $(i=1,2,...,h)$ is $k(i-1)+1,k(i-1)+2,...,ki$. $H_2,..., H_J$ is constructed by randomly arranging each column of $H_1$. Fig. \ref{CheckMatrixAndTannerDiagram} shows a check matrix of (12,3,6) LDPC codes constructed by a Gallager code.

After getting the check matrix $H$, if each row of the matrix is linearly independent, the generator matrix $G$ can be obtained through Gaussian elimination. The steps of the Gaussian elimination are to convert the check matrix H into the form of a systematic code matrix $H_s=[P|I]$ and then obtain the systematic code generation matrix $G_s=[I|P^T]$. For source information, $u=\left\{ u1,u2,u3,...\right\} $, the codeword is ${c=uG}_s$. The Gaussian elimination generally destroys the sparse characteristic of the original parity check matrix, leading to increased coding complexity, and when the packet length is $n$, the coding complexity is $O(N^2)$. 

In addition to Gaussian elimination, approximate lower triangular matrix coding is also a commonly used method. Under the premise of ensuring that the row weight and column weight are fixed, the algorithm rearranges the rows and columns of the matrix $H$ until it becomes an approximate lower triangular matrix with the form $H=\left[\begin{matrix}A&B&T\\C&D&E\\\end{matrix}\right]$. $A$, $B$, $C$, $D$, and $E$ are sparse matrices, and T is a sparse lower triangular matrix. Multiply the matrix $\left[\begin{matrix}I&0\\-ET^{-1}&I\\\end{matrix}\right]$ on its left to get $H^{\prime}=\left[\begin{matrix}A&B&T\\-ET^{-1}A+C&-ET^{-1}B+D&0\\\end{matrix}\right]$. Assuming the original information is u, the codeword is $c=[u, p_1, p_2]$, where $p_1$ and $p_2$ are the check bits and can be calculated by the formula $H^{\prime} c^T=0$. The complexity of calculating $p_1$ and $p_2$ is $O(N+g^2)$ and $O(N)$ respectively, where $g$ represents the length of $p_1$. Take the regular LDPC(12,3,6) code mentioned above as an example, its check matrix and corresponding Tanner diagram are shown in Fig. \ref{CheckMatrixAndTannerDiagram}. Assuming $u=[101010]$, the codeword will be $c=[101010001000]$.

\begin{figure*}[!t]
	\centering
	\includegraphics[width=5in]{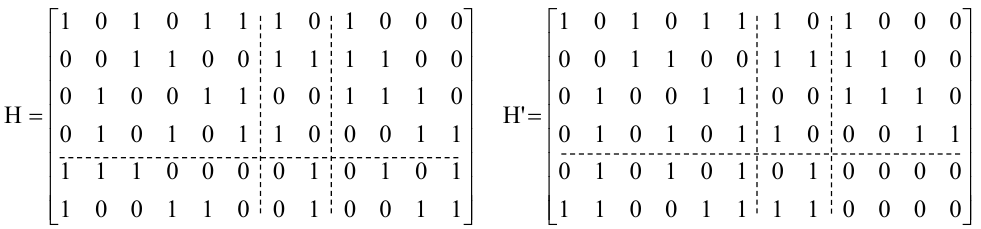}
	\caption{Approximate lower triangular matrix coding of LDPC(12,3,6).}
	\label{ApproximateLowerTriangularMatrixCoding}
\end{figure*}

The decoding algorithms of LDPC codes are divided into two categories: hard-decision decoding algorithms, such as the bit flipping algorithm, and soft-decision decoding algorithms such as the message passing algorithm and the sum-product algorithm. In general, soft-decision decoding algorithms have better performance, while hard-decision decoding algorithms have low computational complexity.

Taking bit flip decoding as an example, the decoding steps are: (1) Compute all parity checksums. Stop decoding when all parity checksums are $0$. (2) For each bit of the received codeword $v$, calculate the error number of the parity check equation related to it, denoted as $f_i, i=0,1,...,n-1$. Select bits whose $f_i$ is greater than the threshold $\delta$ to form a set $S$. (3) Flip the bits in the $S$. (4) Repeat the above steps until all parity checksums are $0$ or the maximum number of iterations is reached. Assuming that the received codeword is $v=[101010001001]$, the numbers of error equations corresponding to these bits are $[201210221223]$. Select the bit with the largest error number to form the set $S$, and then the codeword after flipping is $[101010001000]$, which is the correct codeword we want.

LDPC codes can be combined with watermark codes in DNA storage, which we simulate and compare with other codes in the next section.




\subsection{Polar Code}


In \cite{chen2021artificial}, the watermark codes can correct errors such as insertion and deletion errors, and residual substitution errors are corrected by using LDPC codes. However, the cost of DNA synthesis increases dramatically with the length of the DNA strand \cite{heckel2018archive}. In short code sequences, polar codes perform better than LDPC codes \cite{belhadj2021error}, so we propose to replace the LDPC codes with the polar codes and then concatenate them with watermark codes for error correction.

In 2007, Arikan \cite{arikan2009channel} proposed polar codes as a linear channel coding method. Polar codes convert the original $N$ channels into $M$ channels with capacity tending to 1 and $N-M$ channels with capacity tending to 0 via channel combination and channel decomposition. 

\begin{figure}[!t]
 	\centering
    \subfigure[The smallest compound channel.]{
    	\includegraphics[width=3in]{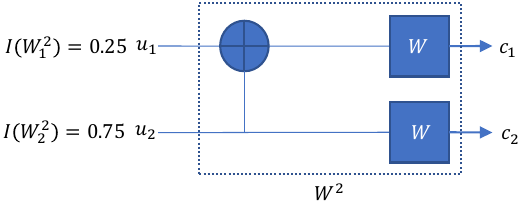}
    	\label{SmallestPolarCodeChannel}
        }
        
    \subfigure[N compound channels.]{
    	\includegraphics[width=3in]{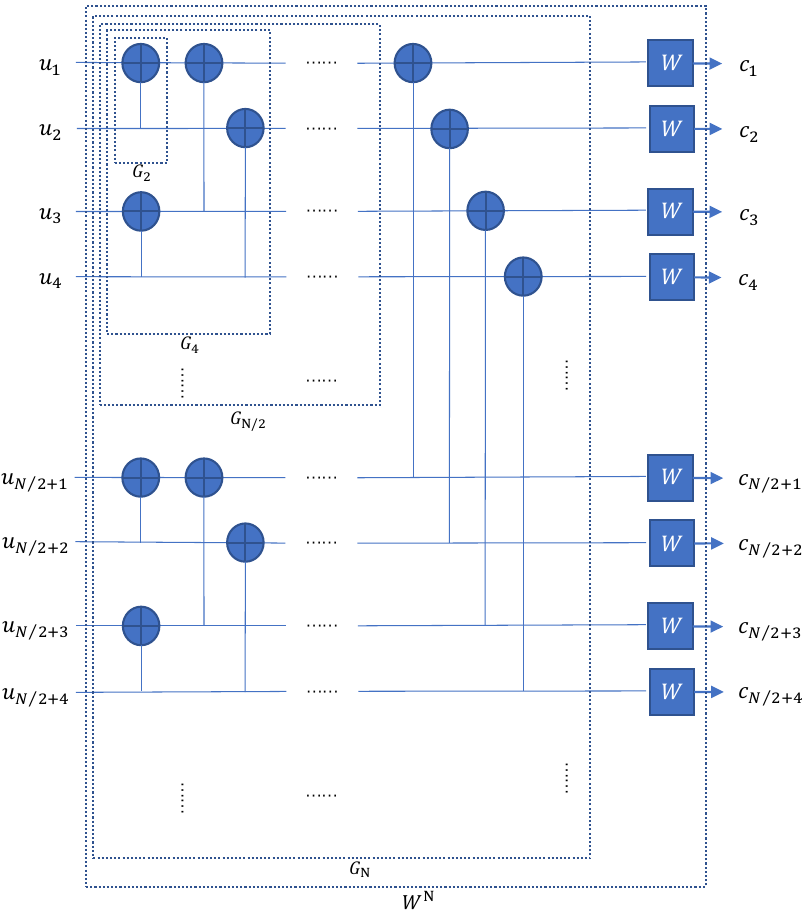}
    	\label{NPolarCodeChannsel}
        }
    \caption{Compound channels for polar codes}
    \label{CompoundChannel}
\end{figure}

Fig. \ref{SmallestPolarCodeChannel} shows the smallest polar code channel $W^{2}$. Assuming the channel $W$ is a binary erasure channel with erasure probability $p$, the capacity $I(W)$ is $1-p$. In $W^{2}$, $c_{1}$ is encoded by $u_{1}+u_{2}$, and $c_{2}$ is $u_{2}$. When we decode the value of $u_1$, we can get the value of $u_1$ only if both $c_1$ and $c_2$ have successfully transmitted. Hence, the capacity of the channel $W_{1}^{2}$ where the bit $u_{1}$ is located can be expressed as
\begin{equation}
I(W_1^2)=(1-p)^2.
\end{equation}
The successful decoding of $u_{2}$ in the channel $W_{2}^{2}$ requires the successful transmission of either $c_1$ when the value of $u_1$ is known or $c_{2}$. Hence, we can get 
\begin{equation}
I(W_2^2)=1-p^2.
\end{equation}

Assuming $p$ is 0.5, we have $I(W_1^2)$ = 0.25 and $I(W_2^2)$ = 0.75, indicating that the polarisation of $I(W_1^2)$ tends to zero and the polarisation of $I(W_2^2)$ tends to 1. We transmit information bits on the channel whose capacity approaches 1, and we transmit 0 bits, also called frozen bits, on the channel whose capacity approaches 0. That is, when we perform $(N,k)$ polar coding, we choose $k$ sub-channels with the highest reliability to transmit information bits, and $N-k$ sub-channels with lower reliability to transmit frozen bits.

We now detail the polar encoding process.
The generator matrix $G_n$ for polar codes is a $n$th Kronecker product of $G_2$:
\begin{equation}
\begin{array}{l}
G_{N}=G_{2}^{\otimes n}=\left[\begin{array}{lc}
G_{N / 2} & 0 \\
G_{N / 2} & G_{N / 2}
\end{array}\right], \\
G_{2}=\left[\begin{array}{ll}
1 & 0 \\
1 & 1
\end{array}\right].
\end{array}
\end{equation}
Hence, we can get the polar code encoded sequence:
\begin{equation}
c_{N}=u_{N} G_{N}.
\end{equation}

\begin{figure}[!t]
	\centering
	\includegraphics[width=0.48\textwidth]{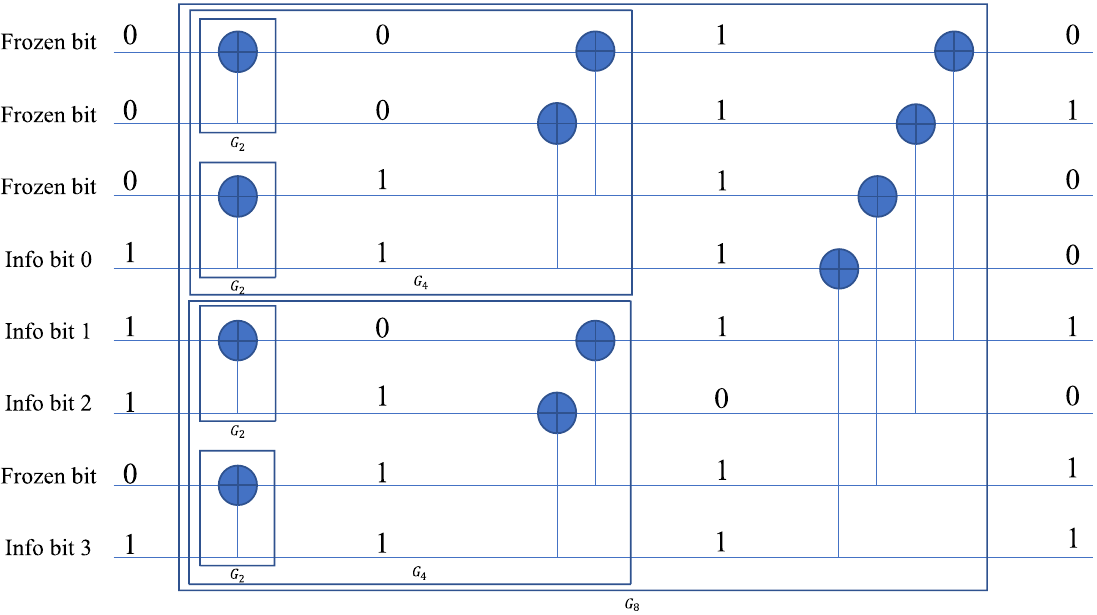}
	\caption{An example of polar encoding.}
	\label{ExampleOfPolarEncoding}
\end{figure}

The encoding process of polar codes is exemplified in Fig. \ref{ExampleOfPolarEncoding}.
Assuming that all information bits are 1 and all frozen bits are 0, the sequence is [0 0 0 1 1 1 0 1]. The generator matrix $G_8$ is
\begin{equation}
G_{8}=\left[\begin{array}{llllllll}
1 & 0 & 0 & 0 & 0 & 0 & 0 & 0 \\
1 & 1 & 0 & 0 & 0 & 0 & 0 & 0 \\
1 & 0 & 1 & 0 & 0 & 0 & 0 & 0 \\
1 & 1 & 1 & 1 & 0 & 0 & 0 & 0 \\
1 & 0 & 0 & 0 & 1 & 0 & 0 & 0 \\
1 & 1 & 0 & 0 & 1 & 1 & 0 & 0 \\
1 & 0 & 1 & 0 & 1 & 0 & 1 & 0 \\
1 & 1 & 1 & 1 & 1 & 1 & 1 & 1
\end{array}\right].
\end{equation}
After polar encoding following Fig. \ref{ExampleOfPolarEncoding}, the encoded sequence becomes [0 1 0 0 1 0 1 1].

\begin{figure}[!t]
	\centering
	\includegraphics[width=0.48\textwidth]{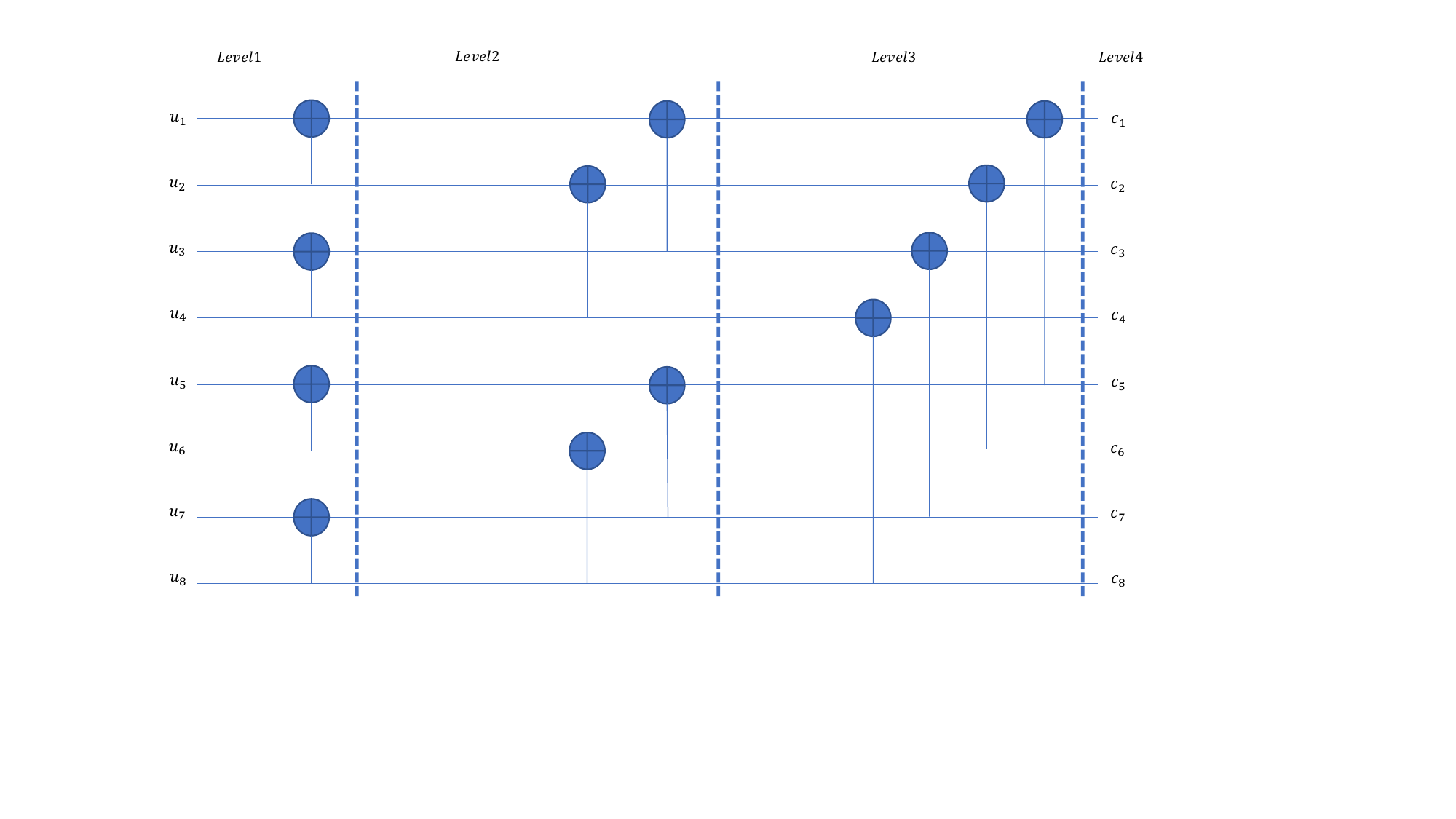}
	\caption{An example of polar decoding.}
	\label{ExampleOfPolarDecoding}
\end{figure}

\begin{figure}[!t]
 	\centering
    \subfigure[LLRs propagate]{
    	\includegraphics[width=3in]{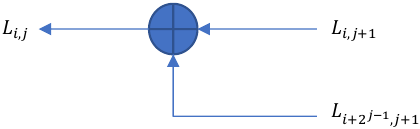}
            \label{LLRsPropagate}
    	}
    \subfigure[Propagate from LLRs to bits]{
    	\includegraphics[width=3in]{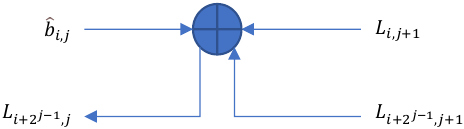}
            \label{PropagateFromLLRsToBits}
            }
    \subfigure[Bits propagate]{
    	\includegraphics[width=3in]{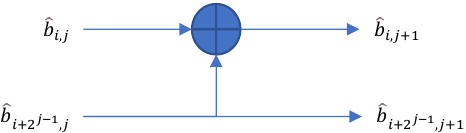}
            \label{BitsPropagate}
            }
    \caption{Three types for SC decoding.}
    \label{ThreePolarCodeDecodingFormulaTypes}
\end{figure}

Contrary to the polar code encoding performed from left to right, the decoding of the polar code is from right to left. Here we take the simplest successive cancellation (SC) decoding algorithm \cite{arikan2009channel} as an example.
We divide the decoding process $Level1, \ldots, Level(\text{log}_2(N)), Level(\text{log}_2(N))+1$ into different levels through the XOR processes \cite{babar2019polar} in Fig. \ref{ExampleOfPolarDecoding}, where $Level1$ represents the decoded bits and $Level(\text{log}_2(N))+1$ represents the received likelihood information. There are three types of SC decoding, as shown in Fig. \ref{ThreePolarCodeDecodingFormulaTypes}, where $L$ and $\hat{b}$ denote soft and hard information respectively, $i$ denotes the $i$-th bit, and j denotes the level of $L$ or $\hat{b}$. \textcolor{black}{More details about the calculation methods of the three types can be found in \cite{babar2019polar}.}


There are also sophisticated polar decoding algorithms, such as 
successive cancellation list decoding \cite{tal2015list}, belief propagation decoding \cite{arikan2008performance} and soft cancellation decoding \cite{fayyaz2014low}.




The construction of the polar code is a  channel selection problem. On the basis of the existing polar code research, it is necessary to obtain the relevant parameters of the DNA storage channel to determine the reliability parameters in the polar code. When the polar code and the watermark code are cascaded for DNA storage error correction, the B-DMC-based density evolution method should be used to estimate the polar code parameters. Because the decoding performance of the watermark code is only related to the error rate of the channel in Fig. \ref{Watermark_Simulation}, when the polar code is used for channel estimation, it can be directly calculated from the decoding result of the watermark code.



 

\begin{figure}[!t]
 	\centering
    \subfigure[Nanopore sequencing model]{
    	\includegraphics[width=3in]{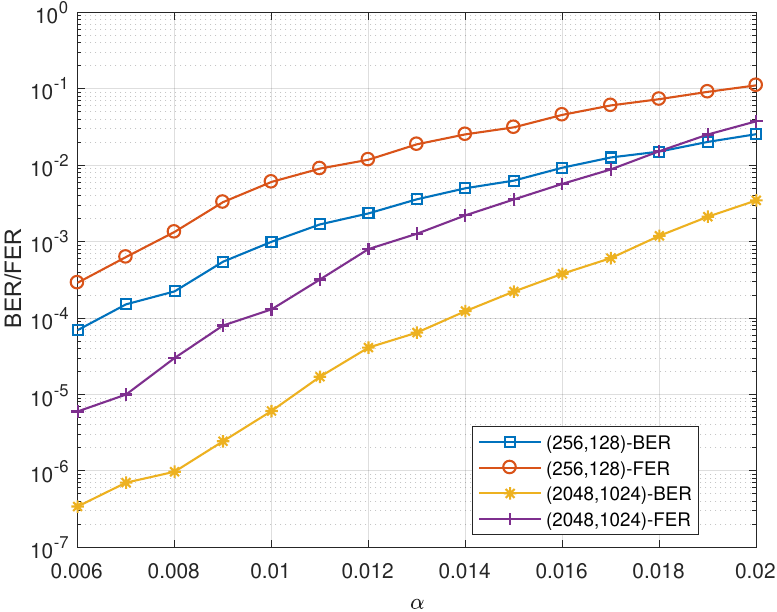}
    	}
    \subfigure[Illumina sequencing model]{
    	\includegraphics[width=3in]{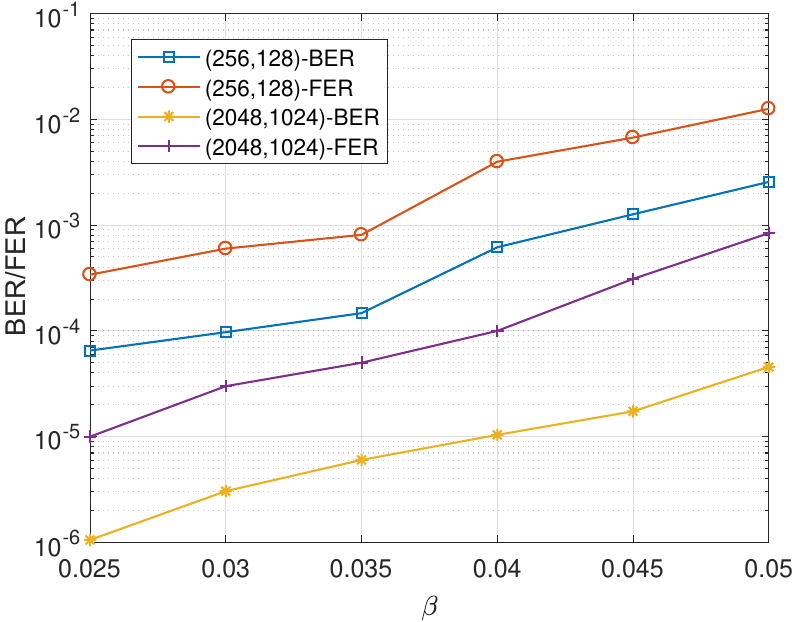}
        }
    \caption{Error correction performance of the polar code.}
    \label{PolarCodeInTwoSequencingModel}
\end{figure}

Given an assumption of exclusive substitution errors within the channel, Fig. \ref{PolarCodeInTwoSequencingModel} depicts both the BER and FER performances of the polar code in two distinct sequencing channels: Nanopore and Illumina. Here, parameters are such that $\alpha \in [0.006, 0.02]$, $\beta$ resides in the interval [0.025, 0.05], and the code rate of the polar code is set at $0.5$.  It is observed that maintaining a constant code rate, and an extended code length enhances error correction efficacy across both sequencing platforms.

\begin{figure}[!t]
 	\centering
    \includegraphics[width=3in]{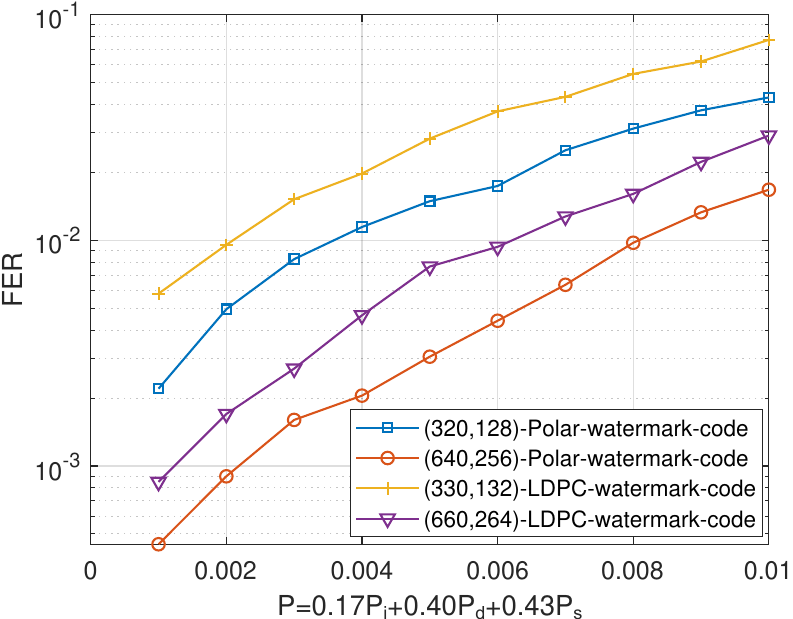}
    \caption{Performance of the polar and LDPC code concatenated with the watermark code.}
	\label{Polarwatermark}
\end{figure}

When there are insertion, deletion, and substitution errors in the channel, we concatenate the outer polar code and the inner watermark code. The corresponding FER is shown in Fig. \ref{Polarwatermark}. The FER of the LDPC-watermark concatenated code is also shown as the benchmark.
The insertion, deletion, and substitution probabilities account for 17$\%$, 40$\%$ and 43$\%$ of the total error probability, respectively. In the simulation, the inner watermark code provides soft information for the polar decoder or LDPC decoder. The polar code uses the SCL decoding algorithm. The complexity of the successive cancellation list decoding (SCL) is $O(LN \log N)$, where $L$ represents the number of most likely decoding paths. The LDPC code uses QC-LDPC code and adopts the MS algorithm with 30 iterations during decoding. 
\textcolor{black}{As shown in Fig. \ref{Polarwatermark}, the FER of polar-watermark code is lower than LDPC-watermark code in the same sequence, hence, the decoding performance of polar-watermark code is better than that of LDPC-watermark code.}

\section{Bottleneck and Future Work}

The development of DNA coding goes hand in hand with DNA storage and communication. At present, there are still many factors hindering their large-scale application. The high cost of DNA synthesis and sequencing is one of the major limitations, and the existing data read latency still maintains a high level. The current record for DNA digital data storage is still around 200 MB, with single synthesis runs lasting about 24 h \cite{doricchi2022emerging}. Technologies like portable sequencing and fully automated data storage systems are expected to drive the commercialization of DNA data encoding \cite{yazdi2017portable,takahashi2019demonstration}. One of the advantages of DNA over traditional data storage is that it can be stored for long periods of time, but only under the proper storage conditions. However, methods like cryopreservation are expensive and cumbersome, and widespread adoption of DNA storage requires research on how to maximize the lifespan of DNA in a dry state at room temperature. In addition, how to quickly locate the required content and modify DNA strands is also challenging. Despite the relevant research progress \cite{lin2020dynamic,lee2020dna}, the amount of DNA data supporting modification is not high, and this is also costly and time-consuming. At this stage, the performance of DNA storage is still far behind traditional storage media such as disks, and it is not suitable for storing data that requires frequent access and modification.

For DNA communication, it has higher requirements on time delay, and additional consideration needs to be given to the transmission of DNA molecules. Considering that DNA molecules are usually released in a static liquid environment, a fluid environment, or transported by a carrier, it is difficult to achieve fast, directional, and long-distance DNA molecule transmission. Many ideal DNA communication application scenarios are at the nanoscale, and it is difficult to fabricate nanomachines with the functions of synthesizing, releasing, and receiving DNA. Due to the above constraints, the DNA communication system is still mainly in the theoretical research stage, while the DNA storage system has already undergone large-scale experiments.

In terms of DNA coding, the compatibility and adaptability of many schemes are limited due to the lack of in-depth research on the properties of DNA molecules. Most of the existing DNA storage coding algorithms only make small adaptive modifications on the basis of traditional error correction coding. Carrying out more experiments related to DNA encoding or using some DNA coding simulation platforms \cite{yuan2022desp,jiang2023dna} will help establish a more comprehensive model of the whole process of DNA storage. Future work should consider more DNA characteristics and biotechnological characteristics, and on this basis, form a more DNA-friendly coding method. When designing codes, multiple copies generated during DNA synthesis and sequencing can also be used for error correction to reduce redundancy from additional error-correcting codes. \textcolor{black}{In addition, most DNA storage systems design data compression, error correction coding, and base coding separately, without considering issues such as the full utilization of resources. Therefore, joint source-channel coding (JSCC) may achieve higher gains. In recent years, machine learning has achieved impressive results in many fields, and it has the potential to be applied in DNA-based JSCC. On this basis, semantic communication can also be introduced into DNA storage, so as to store the semantics of source information more accurately and increase data density.}
 
Furthermore, both DNA-based data storage and communication require the process of encoding. In DNA storage systems, DNA can be eroded and damaged by physical or chemical factors, requiring coding to correct errors. In a  DNA molecular communication 
system, channel coding is required to deal with errors caused by base-level interference (BLI), sequence-level interference (SLI), etc. during transmission. However, much of the current work on DNA-based MCS has been done separately. There have been studies aware of the importance of combining DNA-based data storage with molecular communication systems \cite{shah2017molecular}, but no joint design has been provided either. Given the strong correlation between the two, incorporating DNA-based data storage into the DNA-based communication system has aroused great interest.

\bibliography{references }
\nocite{*}


 





\end{document}